\documentclass[11pt]{article} 
\usepackage[utf8]{inputenc}
\usepackage{multirow}
\usepackage{amsfonts}
\usepackage{amsmath}
\usepackage{amssymb}
\usepackage{amsthm}
\usepackage{caption}
\usepackage[dvipsnames]{xcolor}
    \definecolor{darkgreen}{rgb}{0,0.5,0}
    \definecolor{darkblue}{rgb}{0,0,0.6}
    \definecolor{purple}{rgb}{0.4,.2,0.7}
\usepackage[margin = 2.5 cm]{geometry}
\usepackage{graphicx}
\usepackage[hyperfootnotes = false, colorlinks = true, linkcolor = darkblue, citecolor = purple]{hyperref}
\usepackage{subcaption}
\usepackage{ytableau}

\usepackage{cite}

\usepackage{comment}

\newcommand{\be}{\begin{equation}}
\newcommand{\ee}{\end{equation}}
\newcommand{\bea}{\begin{eqnarray}}
\newcommand{\eea}{\end{eqnarray}}
\def\la{\label}
\def\nref#1{(\ref{#1})}

\newcommand{\Eb}{E_{\text{brk}}}

	\newcommand{\bes}{\begin{equation} \begin{split} }	
	\newcommand{\ees}{\end{split} \end{equation} }

	\newcommand{\lp}{\left (}
	\newcommand{\rp}{\right )}
	\newcommand{\lb}{\left [}
	\newcommand{\rb}{\right ]}
	\newcommand{\RA}{\Rightarrow}

	\newcommand{\Z}{\mathbb{Z}}

	\newcommand{\e}{\varepsilon}

	\usepackage{physics}

\begin{document}

	\thispagestyle{empty}
\begin{center}
    ~\vspace{5mm}

  {\LARGE \bf {Following the state of an evaporating charged black hole into the quantum gravity regime 
  \\}} 

   \vspace{0.5in}
     
   {\bf    Anna Biggs$^1$
   }

    \vspace{0.5in}

  $^1$
  Jadwin Hall, Princeton University,  Princeton, NJ 08540, USA 
   \\
                
    \vspace{0.5in}

    \vspace{0.5in}

\end{center}

\vspace{0.5in}

\begin{abstract}

We study the energy probability density function of an evaporating near-extremal charged
black hole. At sufficiently low energies, such black holes experience large quantum metric fluctuations
in the $AdS_{2}$ throat which are governed by a Schwarzian action. These fluctuations modify
Hawking evaporation rates, and therefore also affect how the black hole state evolves over time.
In previous work on Schwarzian-corrected Hawking radiation, the black hole was taken to be in
the microcanonical or canonical ensemble \cite{Brown:2024ajk}. However, we find that an initially fixed-energy
or fixed-temperature state does not remain so in the regime where Schwarzian corrections are
important. We consider three decay channels: the emission of massless scalars, photons, and
entangled pairs of photons in angular momentum singlet states. In each of the three cases, we find that
in the very low energy,  quantum dominated  regime, the probability distribution of the black hole energy level occupation
tends toward  a particular attractor function that effectively depends on only one combination of time and energy.  This function is independent of the initial state and gives new predictions for the energy fluxes and Hawking emission
spectra of near-extremal charged black holes.

 \end{abstract}
 
\vspace{1in}

\pagebreak

\setcounter{tocdepth}{3}
{\hypersetup{linkcolor=black}\tableofcontents}

			\section{Introduction}

It has been understood since the 90s that the semiclassical treatment of black hole thermodynamics breaks down in the extremal limit \cite{Preskill:1991tb, Maldacena:1998uz}. In short, the near-horizon metric has  soft modes whose action becomes unsuppressed at low temperatures. In this paper our focus will be on 4d Reissner-Nordström black holes. Letting $E \equiv M - Q$ denote the energy of such a black hole above extremality, quantum metric fluctuations begin to modify the semiclassical description at the scale 
	\begin{align}\la{Ebrk}
		E_{\text{breakdown}} \equiv  \frac{\pi }{r_{+}S_{0}}
	\end{align} 
	The quantum gravity effects are parametrically large in $E_{\text{brk}}/E$ or $\beta E_{\text{brk}}$, so the usual perturbative treatment of quantum fields in a fixed background is not valid below this cutoff. 

However, studies in JT gravity from the last decade have revealed how to  perform the gravitational path integral over these ``Schwarzian'' modes, so named because they have an effective description in which their action is a Schwarzian time derivative. As the black hole approaches extremality, the proper distance to the horizon becomes increasingly long, and this region develops a  throat-like spatial geometry with the approximate metric $AdS_{2} \times S^{2}$. The Schwarzian mode corresponds to time-dependent fluctuations in the length of this throat. Performing a dimensional reduction to $AdS_{2}$ in the throat produces a theory of 2d gravity coupled to matter (JT gravity) which is simple enough to quantize. Within this context, it was understood how the Schwarzian modes modify the black hole density of states and thermodynamics below $E_{\text{brk}}$ \cite{Iliesiu:2020qvm, Ghosh:2019rcj, Iliesiu:2022onk, Kapec:2023ruw, Rakic:2023vhv, Kolanowski:2024zrq, Moitra:2019bub, Modak:2025gvp, Boruch:2022tno, Heydeman:2020hhw}, see \cite{Mertens:2022irh} for a review. Recent papers have further investigated Schwarzian corrections to the black hole emission rates and absorption cross section \cite{Brown:2024ajk, Maulik:2025hax, Emparan:2025sao, Bai:2023hpd}.

In this paper we study the probability density function of the black hole energy and its dynamics under Schwarzian-corrected Hawking radiation. Previous analyses of Hawking spectrum near extremality have taken the black hole to be in the canonical or microcanonical ensemble. However, we find that these are not states that the black hole would occupy below $\Eb$ at long times. There,  the probability density of energy level occupation tends toward a particular, non-thermal  distribution which is independent of the initial state. We evaluate the energy flux and particle emission spectra in this state to find new predictions for these observables which differ from those obtained using a microcanonical or canonical state.

First we will consider the evolution under the emission of neutral, massless scalar particles. Although our universe is not known to contain a massless scalar, this case is mathematically simplest and contains most of the interesting details of the higher spin cases.

	The emission processes we consider next are guided by  \cite{Brown:2024ajk}, where the authors study the evaporation history of a charged black hole coupled to the Standard Model. Provided that the black hole has a sufficiently large initial charge,\footnote{Here ``sufficiently large'' means a charge greater than $\sim  1.8 \times 10^{44}q$  where $q$ is the positron charge. Above this charge, Schwinger pair production is exponentially suppressed and the black hole will lose energy faster than charge, driving it toward extremality.} the authors show that it will spend the majority of its lifetime at energies $E \lesssim E_{\text{brk}}$ where Schwarzian corrections are important. They explain that in this regime, the black hole alternates between one of two dominant radiation channels: emission of single photons from a black hole with angular momentum $j = 1/2$ or emission of entangled pairs of photons with zero net angular momentum  (``di-photons'') from a black hole with $j = 0$. This motivates us to study the evolution of the black hole state under these two radiation channels in the Schwarzian regime.

In all cases, we find an altered Hawking spectrum with a larger energy flux than the one in the microcanonical ensemble. The expected value of the black hole energy $\langle E(t)\rangle$ has the same power-law time dependence as in the microcanonical ensemble, but with a prefactor that is as much as $\sim 700$ times larger. In particular, we find that the time dependence of the expected energy is fixed by a scaling symmetry of the long-time solution. This scaling symmetry is different than the conformal symmetry present in the semiclassical regime. 

The rest of the paper is organized as follows. In order to connect 4d observables to 2d JT gravity, we use a low-energy effective description of the black hole's interactions with the probe field. We review this effective description in section \ref{ET}, illustrating the ideas for the case of a neutral massless scalar. 	In section \ref{PEa1} we discuss the equation governing the black hole probability distribution of energy level occupation and its solution in the semiclassical regime. We then solve the equation at energies below $E_{\text{brk}}$ in the scalar case and discuss the solution's long-time behavior. In sections \ref{PEa3} and \ref{PEa7} we perform a similar analysis for the evolution under photon and di-photon emission. In section \ref{Spec} we present results for the corrected emission spectra and energy fluxes. 
Finally, in section \ref{ACS} we comment on Schwarzian corrections to the absorption cross section of massless scalars.

While we were preparing this paper, we learned of work by Roberto Emparan \cite{Emparan:2025sao} which has some overlap with our discussion in section \ref{ACS}.

	\section{Low energy effective theory}\la{ET}
	
	In this section, we consider a neutral, massless scalar field coupled to a near-extremal Reissner-Nordström black hole. We review how interactions with the black hole are captured by an effective theory which replaces the black hole by a non-gravitating quantum system. This effective description has been employed explicitly and implicitly in many previous papers. To name just a few examples, see \cite{Brown:2024ajk, Emparan:2025sao, Maulik:2025hax, Callan:1996tv, Das:1996wn, Gubser:1996xe, Gubser:1996zp, Gubser:1997cm, Maldacena:1996ix, Goldberger:2005cd, Goldberger:2019sya, Biggs:2024dgp, Bai:2023hpd}. Readers familiar with this description can safely skip to section \ref{PEa1}.

	The idea is that, for the purpose of describing low-energy interactions with the probe scalar from many Schwarzschild radii away, we can model the black hole by a quantum system living at a point in Minkowski spacetime. In particular, the point particle approximation is valid in the low frequency limit $r_{+} \omega \ll 1$ where the probe does not resolve the finite size of the black hole. 
	The coupling between the 4d scalar $\phi$ and the black hole is captured by a Hamiltonian of the form
	\begin{align}\la{Hint}
		H_{\text{int}} = g O(t) \phi(t,\vec{0})~~~~~~~~~~ r_{+}\omega \ll 1
	\end{align}
	Here we have chosen to work in the frame where the black hole sits stationary at the origin, where it interacts with $\phi$. $O$ is an operator living on the point particle worldline which acts on the black hole Hilbert space. In the language of AdS/CFT, it is the $\Delta = 1$ primary operator dual to the massless scalar in $AdS_{2}$.\footnote{$O$ can alternatively be viewed as living at the boundary of the $AdS_{2}$ near-horizon region, where it couples to the boundary value of the 2d scalar field. The two presentations are equivalent. The effective description of the black hole as a point particle sometimes goes under the name  ``worldline effective theory'' in the gravitational waves literature.} $g$ is a coupling which we will determine shortly.

	We will ignore changes in the black hole center of mass momentum due to emission and absorption. This effect would lead to corrections scaling as $\omega/M \sim r_{+} \omega/S_{0} \ll 1$ which are subleading in the low-frequency limit we consider. In principle, $H_{\text{int}}$ contains additional operators which couple to derivatives of the scalar field, but those interactions are suppressed by additional powers of $r_{+}\omega$ relative to \nref{Hint}.
	
	It will be important for our analysis that	\nref{Hint} is weakly coupled, allowing us to do perturbation theory in $H_{\text{int}}$. One way to see this is to note that $[O] = 1$ and $[\phi]=1$, so $\lb g\rb=-1$, meaning that $g$ is an irrelevant coupling and therefore weak for the IR scattering processes of interest.

	Another way is by solving the wave equation in the black hole background. The scalar field feels a gravitational potential barrier which separates the near-horizon $AdS_{2} \times S^{2}$ region of the geometry from the asymptotically flat region. The transmission probability for scattering through the potential barrier scales as $|T|^{2} \propto (r_{+}\omega)^{2}$ to leading order in $r_{+}\omega$. So, in the low frequency limit, only a small fraction of the field penetrates the potential barrier to interact with the strongly gravitating region; i.e. the two systems are weakly coupled.

	We determine $g$ by a matching computation. Specifically, we calculate a 4d observable of the system, such as the emission rate or absorption cross section of the black hole,  first using the effective theory \nref{Hint} and second using standard perturbative QFT in the fixed Reissner-Nordström background. The former boils down to an expression involving the $\langle OO \rangle$ two point function, which in the semiclassical limit---where the standard perturbative result is valid---is fixed by $AdS_{2}$ isometries to be the conformal correlator of a $\Delta = 1$ operator. Equating the two results fixes $g$. 
	
Matter correlators such as the $\langle OO \rangle $ two point function have been computed within the context of JT gravity including the gravitational path integral over the Schwarzian mode\footnote{Here we consider only the path integral over genus zero topologies. Corrections to the two point function due to wormhole geometries become important at energies of order $E \sim E_{\text{brk}} e^{-S_{0}}$. } \cite{Mertens:2017mtv, Mertens:2018fds, Lam:2018pvp, Blommaert:2018oro, Iliesiu:2019xuh, Kitaev:2018wpr, Yang:2018gdb, Suh:2020lco}.  Once we have expressed an observable in terms of $\langle OO\rangle$, those  previous results immediately tell us how it is modified by the Schwarzian, simply by expanding $\langle OO \rangle$ at low energies $E \ll E_{\text{brk}}$. In the remainder of this section we detail the steps of this calculation.

	 \begin{center}
    \includegraphics[width=0.9\textwidth]{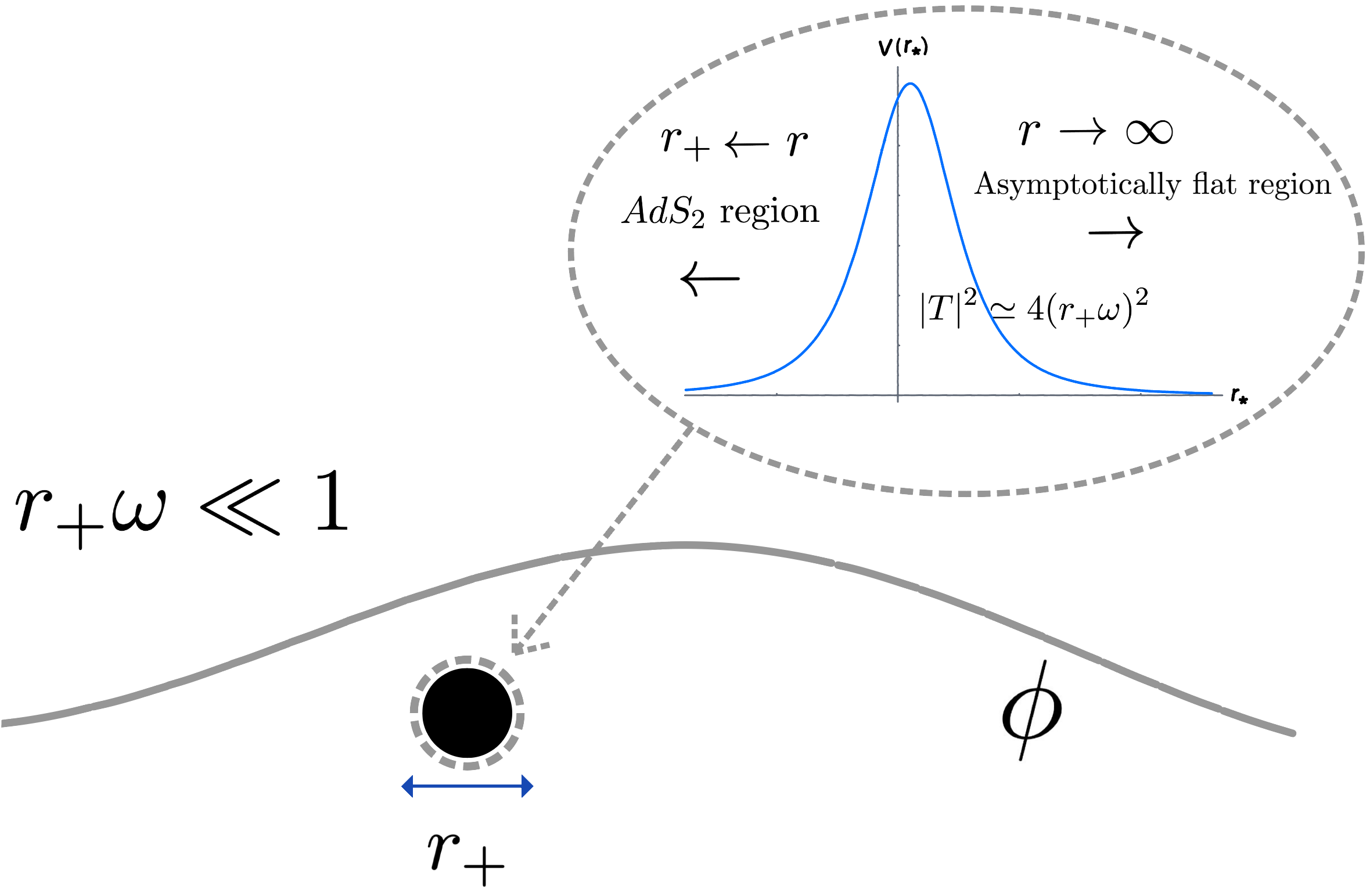}
     \captionof{figure}{A schematic of the effective theory description. We have a probe scalar $\phi$ with wavelength much larger than $r_{+}$, the size of the black hole. The operator $O$ captures the physics of $\phi$ in the strongly gravitating $AdS_{2}$ region, which is separated from the asymptotically flat region by a gravitational potential barrier $V$. These two regions are weakly coupled because the transmission probability through $V$ is small in the low frequency limit.}
 	  \end{center}

	\subsection{Derivation of emission rate in the effective theory}\la{ETemission}
	
	Here we calculate the emission rate due to spontaneous emission of scalar particles using the effective interaction \nref{Hint}.

The Hilbert space is a tensor product of the the black hole and matter hilbert spaces, $\mathcal{H} = \mathcal{H}_{\text{BH}} \otimes \mathcal{H}_{\text{matter}}$. 	The initial and final states are $|i \rangle = |\psi_{i}, 0\rangle $ and $|f\rangle = |\psi_{f}, \vec{q}\rangle$, where $\vec{q}$ is the 3-momentum of the emitted particle and $|\psi_{i,f}\rangle$ are the initial and final states of the black hole. The amplitude for transition from $|i\rangle$ to $|f\rangle$ is then
\begin{align}
	A_{i \to f} = -i g \frac{1}{\sqrt{2|\vec{q}|}} \int_{0}^{T} dt \langle \psi_{f}| O(t) |\psi_{i} \rangle e^{i |\vec{q}|t}
\end{align}
where we have assumed the interaction takes place over a total time $T$. To find the total emission probability over $T$, we square the amplitude and sum over final states. This includes a sum over the final state of the black hole $|\psi_{f}\rangle$ as well as an integral over the 3-momenta $\vec{q}$ of the emitted particle. We also average over the initial states of the black hole if there are many. For example, if the black hole is in a thermal state, this average produces a thermal two point function. This yields
\begin{align}
\Gamma(T) \equiv 	\sum_{E_{i}}p(E_{i}) \sum_{|\psi_{f}\rangle}\int \frac{d^{3}q}{(2 \pi)^{3}} |A_{i \to f}|^{2}
	&=g^{2}T\int_{0}^{\infty}\frac{d\omega }{2 \pi} \frac{\omega^{2}}{\pi}\frac{1}{2 \omega} \int_{-T}^{T} dt ~e^{-i \omega t} \langle  O(t) O(0)  \rangle
\end{align}
where we used time translation invariance of the correlator and relabelled $|\vec{q}| \to \omega$. The number of particles emitted per unit time is then
\begin{align}\la{dNdtet}
	\frac{dN}{dt} = \lim_{T \to \infty}\frac{1}{T} \Gamma(T) = g^{2} \int_{0}^{\infty}\frac{d \omega}{2 \pi}\frac{\omega}{2 \pi}\int_{-\infty}^{\infty}dt e^{-i \omega t} \langle O(t) O(0)\rangle
\end{align}
We reiterate that the meaning of $\langle \rangle$ in \nref{dNdtet} depends on the initial state of the black hole. If the black hole is prepared in a microcanonical state, the operators are evaluated in a fixed energy eigenstate. If the black hole is in a thermal state, the expectation value means $\langle \cdot \rangle = \text{Tr}[e^{-\beta H} ~\cdot ~ ]$. 

\subsubsection{Matching to fix the coupling}
We can now fix $g$ by comparing Hawking's result for the particle flux to \nref{dNdtet} in the semiclassical regime where $E_{\text{brk}} \ll  E$, $ \omega \ll E$. In this limit, the conformal symmetry of the near-horizon geometry is umbroken, and $\langle OO \rangle$ is the finite-temperature conformal correlator of a $\Delta = 1$ operator,
\begin{align}
		\int dt e^{-i \omega t} \langle O(t) O(0) \rangle = \frac{2 \pi \omega}{e^{\beta \omega}-1} \la{confFT}
\end{align}
where
\begin{align}\la{betaE}
	\beta = \sqrt{\frac{2 \pi^{2}}{E_{\text{brk}}E}}
\end{align}
is the inverse Hawking temperature of the black hole.

On the other hand, calculating the emission rate using quantum fields on the fixed 4d black hole background, the rate of neutral scalar particle emission is 
\begin{align}\la{dEdt}
	\frac{dN}{dt} = \int_{0}^{\infty} \frac{d\omega}{2\pi} \frac{\omega }{\pi} \sigma(\omega)\frac{\omega}{e^{\omega \beta }-1}
\end{align}
$\sigma(\omega)$ is the semiclassical absorption cross section. At low frequencies, $\sigma(\omega)$ is famously known to approach the horizon area \cite{Das:1996we},
\begin{align}
	\sigma(\omega) \simeq 4 \pi r_{+}^{2} ~~~~~~~~~ r_{+} \omega \ll 1
\end{align}
Due to the thermal factor $(e^{\omega \beta} -1)^{-1}$, the integral in \nref{dEdt} is dominated by frequencies $r_{+}\omega \ll 1$, so $\sigma(\omega)$ can be approximated by the area. Plugging \nref{confFT} into \nref{dNdtet} and equating it with \nref{dEdt}, we find
\begin{align}\la{g}
	g = 2 r_{+}
\end{align}

We could just as easily have found $g$ by matching the absorption cross section. Schwarzian corrections to the absorption cross section and its expression in the effective theory will be discussed in section  \ref{ACS}.

\subsection{Schwarzian corrections}

In \nref{dNdtet} we expressed the emission rate in terms of $\langle OO \rangle$. 
In the semiclassical regime, this two point function is given by \nref{confFT}. 
The calculation of $\langle OO\rangle$ at energies $E \ll E_{\text{brk}}$ involves integrating over the Schwarzian mode in the JT gravity partition function. For a black hole with zero angular momentum and fixed charge $Q$, the result of this calculation is
\begin{align}
		\int dt e^{-i \omega t} \langle E |  O(t) O(0)|E  \rangle &= 2 \pi \rho(E-\omega) |O_{E,E-\omega}|^{2}\la{Gwmic}
\end{align}
where\footnote{	We adopt the common convention that $\pm$ inside the Gamma function denotes a product over all signs; i.e. $\Gamma(a \pm b) = \Gamma(a + b)\Gamma(a-b)$.}
		\begin{align}
\rho(E) &= \frac{1}{2 \pi^{2}E_{\text{brk}}} e^{S_0}\sinh(2 \pi \sqrt{2 E_{\text{brk}}^{-1} E})\Theta(E)\la{rho}\\
|O_{E_1,E_2}|^{2} &= \frac{2 e^{-S_{0}}\Gamma\lp \Delta \pm i \sqrt{2 E_{\text{brk}}^{-1}E_{1}} \pm i \sqrt{2 E_{\text{brk}}^{-1}E_{2}}\rp }{(2 E_{\text{brk}}^{-1})^{2\Delta}\Gamma(2\Delta)}\la{O}
\end{align}
Here, $\rho(E)$ is the black hole density of states, $S_{0}$ is the Bekenstein-Hawking entropy of the black hole, and $E_{\text{brk}}$ is the energy scale where Schwarzian corrections become important, as discussed around \nref{Ebrk}. We have written the matrix elements for a general dimension $\Delta $ matter operator; for the scalar case under discussion, $\Delta = 1$. The Schwarzian-corrected emission rate is simply given by plugging \nref{rho} and \nref{O} into \nref{dNdtet},
\begin{align}
	\frac{dN}{dt} = \frac{ 2r_{+}^{2}}{ \pi} \int_{0}^{\infty} d\omega \omega  \rho(E-\omega)|O_{E,E-\omega}|^{2}
\end{align}
This formula can be found in \cite{Brown:2024ajk}. In particular, we get the transition rate per unit frequency from a state $|E\rangle$ to $|E - \omega \rangle$ by stripping off the $\omega$ integral over the energy of emitted modes,
\begin{align}\la{scgamma}
	\frac{d^{2}N}{dt d\omega}\bigg|_{E} = \gamma(E,E-\omega)\rho(E-\omega) ~~~~\text{ where } ~~~~\gamma(E,E-\omega)	 = \frac{2 r_{+}^{2}}{\pi} \omega |O_{E,E-\omega }|^{2}
\end{align}
For the remainder of the paper, we will use $\gamma(E,E')$ to denote the microcanonical transition rate from energy eigenstate $|E\rangle$ to $|E'\rangle$. We have already accounted for an overall energy-conserving delta function $\delta(E-E'-\omega)$, which will be left implicit.

	\section{Energy probability distribution - scalar emission}\la{PEa1}
	
	We would like to study the probability distribution of the black hole's energy level occupation as a function of time. Let $P(E,t)$ denote the differential probability of level occupation at energy $E$ and time $t$, and $\gamma(E,E')$ denote the transition rate\footnote{We avoid the traditional notation of $\Gamma$ for the decay rate to avoid confusion with the Gamma function, which will appear frequently.} from $|E \rangle$ to $|E'\rangle$. 	
	This function is governed by an equation of the form
	\begin{align}\la{Peqn}
	\boxed{
		\frac{dP(E,t)}{dt} = -\int_{0}^{E}dE' \rho(E')\gamma(E,E')P(E,t) + \int_{E}^{\infty}dE''\rho(E)\gamma(E'',E)P(E'',t)}
	\end{align}
	
	The change in occupation probability at $E$ comes from two processes. The first, which corresponds to the first term on the righthand side of \nref{Peqn}, is when the black hole decays from $|E\rangle$ to a lower eigenstate $|E'\rangle$. The second is one where a higher eigenstate $|E''\rangle$ decays into $|E\rangle$, captured by the second term in \nref{Peqn}.

It is easy to verify that $P(E,t)$ satisfies probability conservation by integrating \nref{Peqn} with respect to $E$, since the two terms on the righthand side are equal and opposite after we integrate over $E$.

The details of $\gamma(E,E')$ depend on the emission process under consideration. In the following sections we will consider three decay channels: the emission of massless scalars, of $\ell = 1$ photons, and of entangled photon pairs in a singlet state (``di-photons''). The transition rate for scalar emission is \nref{scgamma}. In the low energy limit $E \ll \Eb$ the operator matrix elements become constant, and the transition rate can be approximated as
\begin{align}\la{srate}
\text{ scalar emission } ~~~~~~~~~~~~~~~~\gamma(E,E') &=   \frac{1}{\pi}E_{\text{brk}}^{2}r_{+}^{2}e^{-S_0}(E-E') ~~~~~~~~~~~~~~~~ E \ll \Eb
\end{align}
The other two transition rates (as well as the scalar case) were derived in \cite{Brown:2024ajk}. We quote their results below in the low-energy limit $E \ll \Eb$.
\begin{equation}\la{otherrates}
	\begin{split}
			\text{ $\ell = 1$ photon emission, $j = 1/2$ BH}~~~~~~\gamma(E,E') &= \frac{1}{9 \pi} E_{\text{brk}}^{6}r_{+}^{8}e^{-S_0} (E-E')^{3}~~~~~~~~~~~~~~(*)\\
		\text{di-photon emission, $j = 0$ BH} ~~~~~~~\gamma(E,E') &=(8.2 \times 10^{-4}) \frac{640}{189 \pi^{3}}E_{\text{brk}}^{10}r_{+}^{16}e^{-S_{0}}(E - E')^{7}\\
	 (*) ~~\text{ for }& \frac{3}{8}\Eb  \lesssim E, E'  \ll \Eb 
	\end{split}
\end{equation}

The primary difference between the three cases is in how $\gamma$ scales with the energy of the emitted mode.\footnote{When the black hole has angular momentum $j = 1/2$, the energy spectrum is shifted up, and the density of states goes to zero at $E = \frac{3}{8} \Eb$ rather than at $E = 0$. This is the reason for the additional condition $(*)$. More details about the modified density of states and the role of angular momentum will be given in section \ref{PEa3}.} The powers that appear are 1, 3, and 7, respectively. As we will see, these powers are responsible for the difference in the behavior of $P(E,t)$ between the three cases.

\subsection{Semiclassical regime}

First let us consider  \nref{Peqn} in the semiclassical limit where $E \gg E_{\text{brk}}$ and $\omega \ll E$. We will show that in this regime, \nref{Peqn} is solved by a time-dependent thermal state,
	\begin{align}\la{Pth}
	P_{\text{th}}(E,t) =Z(\beta(t))^{-1}\rho(E)e^{-\beta (t) E}
\end{align}
For non-rotating semiclassical black holes, the differential rate of particle flux per unit frequency is given by
\begin{align}
	\frac{d^{2}N}{dt d\omega} \bigg|_{E} &= \frac{1}{2 \pi}\sum_{\ell,m}\frac{1}{e^{\beta(E) \omega} -1}\mathcal{N}_{\ell,m}\mathcal{P}(\omega,\ell)
\end{align}
Here $\mathcal{P}(\omega,\ell)$ is the greybody factor, also called the absorption probability or transmission coefficient. It represents the proportion of a wave propagating outward from the even horizon that is transmitted through the gravitational potential barrier and escapes to infinity. $\mathcal{N}_{\ell,m}$ is a possible degeneracy factor which counts the modes with quantum numbers $\ell,m$ (e.g. due to multiple polarizations). $\beta(E)$ is the semiclassical inverse Hawking temperature associated with the state of energy $E$, see \nref{betaE}. Here we will only consider the emission of bosonic fields, but if we had fermionic fields, the thermal Plank factor would have a plus sign.

Plugging \nref{Pth} into \nref{Peqn}, we have on the lefthand side
\begin{align}\la{LHS}
	\frac{d P_{\text{th}}(E,t)}{dt} = P_{\text{th}}(E)\dot \beta \lp \langle E \rangle -E\rp 
\end{align}
On the righthand side, we approximate $P_{\text{th}}(E + \omega)$ as
\begin{align}
	P_{\text{th}}(E+\omega) &\approx P_{\text{th}}(E)e^{(\beta(E) - \beta)\omega}\\
	&\approx P_{\text{th}}(E)  \lp 1+ (E - \langle E \rangle) \frac{\partial \beta}{\partial E}\bigg|_{\langle E \rangle} \omega \rp 
\end{align}
where we have expanded $\beta(E)$ around $\langle E \rangle$. We then have, in the $\omega \ll E$ limit,
\begin{align}
\frac{dP_{\text{th}}(E)}{dt}	&=\frac{1}{2 \pi}P_{\text{th}}(E)\sum_{\ell,m} \mathcal{N}_{\ell,m}\int_{0}^{\infty}d\omega \frac{\mathcal{P}(\omega,\ell)}{e^{\beta(E)\omega}-1}\lb -1+1+ (E - \langle E \rangle) \frac{\partial \beta}{\partial E}\bigg|_{\langle E \rangle} \omega\rb \\
	&=-P_{\text{th}}(E)(\langle E \rangle-E) \frac{\partial \beta}{\partial E}\bigg|_{\langle E \rangle}\sum_{\ell,m} \mathcal{N}_{\ell,m}\int_{0}^{\infty}\frac{d\omega}{2 \pi} \omega\frac{ \mathcal{P}(\omega,\ell)}{e^{\beta(E)\omega}-1} \la{RHS}
\end{align}
We recognize in \nref{RHS} the semiclassical expression for the total energy flux,
\begin{align}
	\frac{d\langle E\rangle }{dt} &= -\mathcal{N}_{\ell,m}\int_{0}^{\infty}\frac{d\omega}{2 \pi} \omega\frac{\mathcal{P}(\omega,\ell)}{e^{\beta(E)\omega}-1} 
\end{align}
Assuming $P_{\text{th}}(E)$ is sharply peaked around $\langle E \rangle$, we can evaluate the energy flux at $E = \langle E \rangle$. We then have $\frac{\partial \beta}{\partial E}\frac{d\langle E\rangle }{dt} = \dot \beta$, and \nref{RHS} reduces to \nref{LHS}.

\subsection{Quantum regime}

We now solve \nref{Peqn} for scalar emission in the quantum gravity regime $E \ll E_{\text{brk}}$ where Schwarzian corrections become important. Plugging in \nref{srate} for transition rate, 
the energy probability equation becomes
\begin{align}\la{PQeqn}
	\frac{1}{c_{1}}\frac{dP(E,t)}{dt} = -\int_{0}^{E}dE'\sqrt{E}(E-E') P(E,t) + \int_{E}^{\infty}dE''\sqrt{E''}(E''-E)P(E'',t) \\
	\text{ where } ~~~~~~~c_{1} \equiv \frac{\sqrt{2 E_{\text{brk}} } r_{+}^{2}}{\pi^2}\nonumber ~~~~~~~~~~~~~~~~~~~~~~~~~~~~~~~~~~~~~~~~~~~~~~~~~~~~~~~~~~~~~~~~
\end{align}
\nref{PQeqn} can be solved analytically. The derivation can be found in  Appendix \ref{PQsol}. We find that, subject to the initial condition 
\begin{align}
	P(E, t = 0) = \delta(E - E_{0})
\end{align}
for some $E_{0} \ll E_{\text{brk}}$, the solution for $t>0$ and $E<E_{0}$ is
\begin{align}
	\la{Pfullsol}
	P(E,t) = e^{-\mathcal{E}_{0}\tau}\delta(E-E_{0}) + \frac{3}{2}\frac{1}{E} \lp e^{-\mathcal{E} \tau} - e^{-\mathcal{E}_{0}\tau} \lp \frac{\mathcal{E}}{\mathcal{E}_{0}}\rp^{\frac{3}{5}} -\frac{3}{5}\lp \mathcal{E} \tau\rp^{\frac{3}{5}}\lb \Gamma\lp -\frac{3}{5},\mathcal{E} \tau\rp  - \Gamma\lp -\frac{3}{5}, \mathcal{E}_{0} \tau\rp \rb \rp \nonumber \\
	\text{ for } E<E_{0}, ~~~~~~~~~ \text{ where } ~~~~~~~~~~\mathcal{E} \equiv E^{5/2}, ~~~~~ \mathcal{E}_{0} \equiv E_{0}^{5/2},~~~~~~\tau \equiv \frac{4}{15} c_{1} t~~~~~~~~~~~~~~~~~~~~~~~~~~~~~~~~~
		\end{align}
$\Gamma(a,x)$ denotes the incomplete gamma function $\Gamma(a,x) = \int_{x}^{\infty}\frac{dt}{t}t^{-a}e^{-t}$.  For $E>E_{0}$, $P = 0$ because the black hole is only allowed to decay to a lower energy state.

\subsubsection{Attractor solution}
We can wonder how this distribution behaves at long times. As $t$ increases, the expected energy $\langle E \rangle$ of the black hole will decrease. Therefore, at large times such that $\mathcal{E}_{0} \tau \gg 1$, we still expect that for the energies $E$ where $P$ has support, $\mathcal{E} \tau \sim \mathcal{O}(1)$. This motivates us to consider the limit $\mathcal{E}_{0} \tau \to \infty$ with $\mathcal{E} \tau$ fixed. In this limit, all the terms of \nref{Pfullsol} which depend on $E_{0}$ are exponentially suppressed, and $P(E,t)$ reduces to the simple function
\begin{align}\la{Pbar}
	P(E,t) \to \bar P(E,t) \equiv  \frac{3}{2}\frac{1}{E}\lp \mathcal{E} \tau\rp^{3/5}\Gamma\lp \frac{2}{5}, \mathcal{E} \tau\rp~~~~~~\text{ as } ~~~~~\mathcal{E}_{0} \tau \to \infty 
\end{align}
As a sanity check, we note that $\bar P$ satisfies the normalization condition $\int_{0}^{\infty} dE \bar P = 1$.

The time evolution of any other initial distribution $Q(E,t = 0) = P_{0}(E)$ localized around some $E_{0}$ would be obtained by integrating over the delta function solution \nref{Pfullsol}, and therefore will also approach $\bar P$ at sufficiently long times:
\begin{align}
	&Q(E,t = 0) = \int_{0}^{\infty}dE'\delta(E - E')P_{0}(E')\\
	&\RA \quad Q(E,t) \to \int_{0}^{\infty}dE'P_{0}(E')\bar P(E,t) = \bar P(E,t) ~~~ \text{ as } ~~~ \mathcal{E}_{0} \tau \to \infty
\end{align}
This would suggest that the equation \nref{PQeqn} has an attractor solution given by \nref{Pbar}.

We verify that this is indeed the case by solving \nref{PQeqn} numerically. Our numerical method consists of discretizing $E$ and leaving $t$ continuous, which reduces \nref{PQeqn} to a first-order matrix equation of the form $\frac{d}{dt} \vec{P}(t) = A \vec{P}(t)$. Here $\vec{P}(t)$ is an $n$-component vector of time-dependent functions, where $n$ corresponds to the number of discretized $E$ values.

We find that the numerical solution converges to \nref{Pbar} regardless of the initial distribution. In Figure \ref{a1Pplots} we plot the numerical solution for the time evolution of an initial delta function at $E = E_{\text{brk}}$ against the attractor solution $\bar P(E,t)$ at the same times. Visually we can see the former converging to the latter.

To quantify how quickly this convergence occurs, in Figure \ref{a1diff} we plot the difference between the two functions over time,
\begin{align}\la{delP}
	\Delta P(t) \equiv \int_{0}^{\infty}dE |P_{\text{num}}(E,t) - \bar P(E,t) |
\end{align}
It takes about $E_{\text{brk}}^{5/2}c_{1}t \sim 20$ for the difference $\Delta P(t)$ to be less than $1 \%$.

  \begin{center}
    \includegraphics[width=0.9\textwidth]{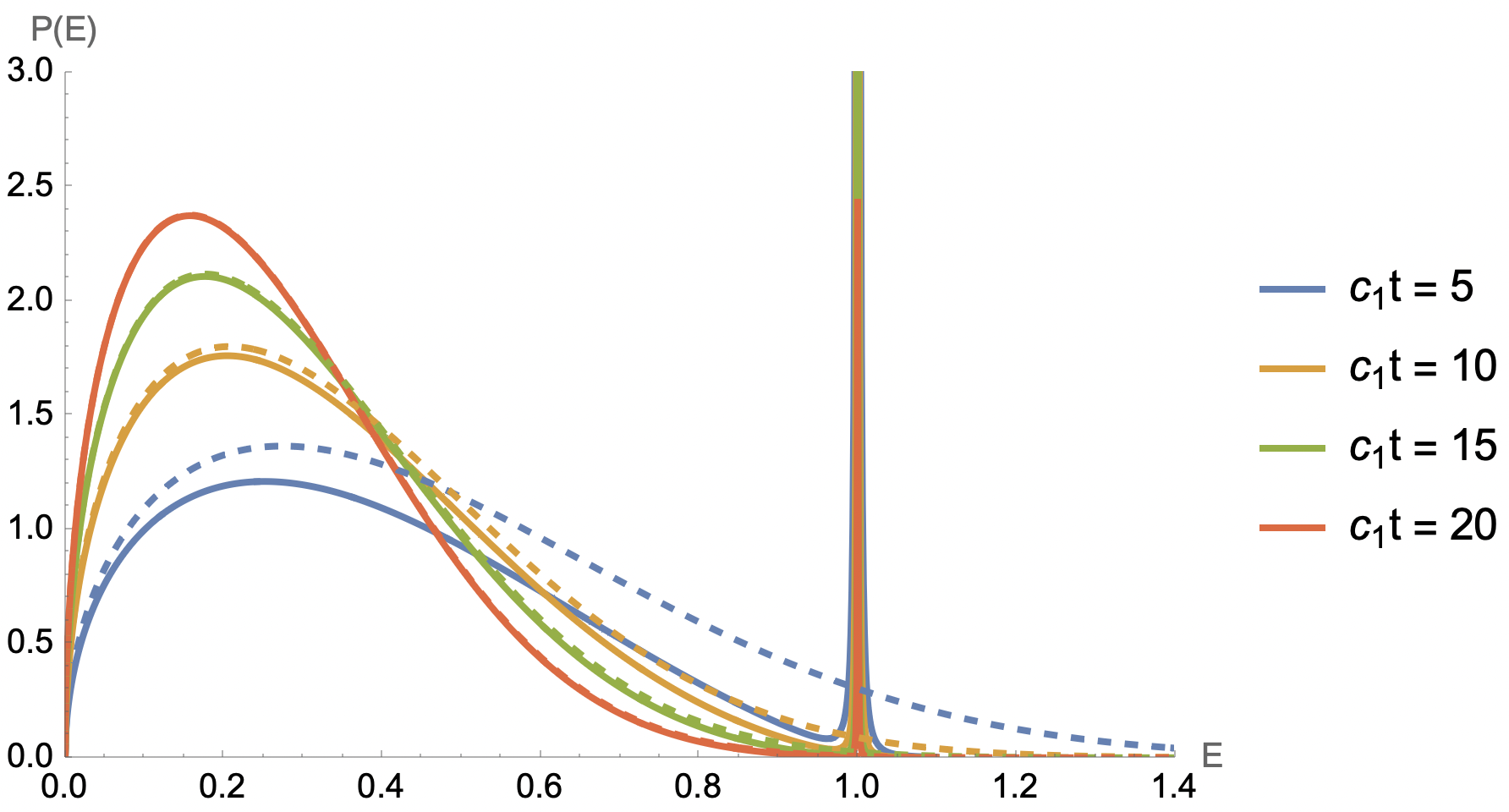}
     \captionof{figure}{Time evolution of $P(E,t)$ for a black hole undergoing scalar emission in the Schwarzian regime. Plots are in units where $E_{\text{brk}} = 1$. Solid lines: the numerical solution to \nref{PQeqn} starting from a delta function distribution at $E_{0} = E_{\text{brk}}$. Dashed lines: the attractor solution $\bar{P}(E)$ \nref{Pbar}. We see the numerical solution converging to the attractor solution over time.}\la{a1Pplots}   
 \end{center}
 
 \begin{center}
    \includegraphics[width=0.7\textwidth]{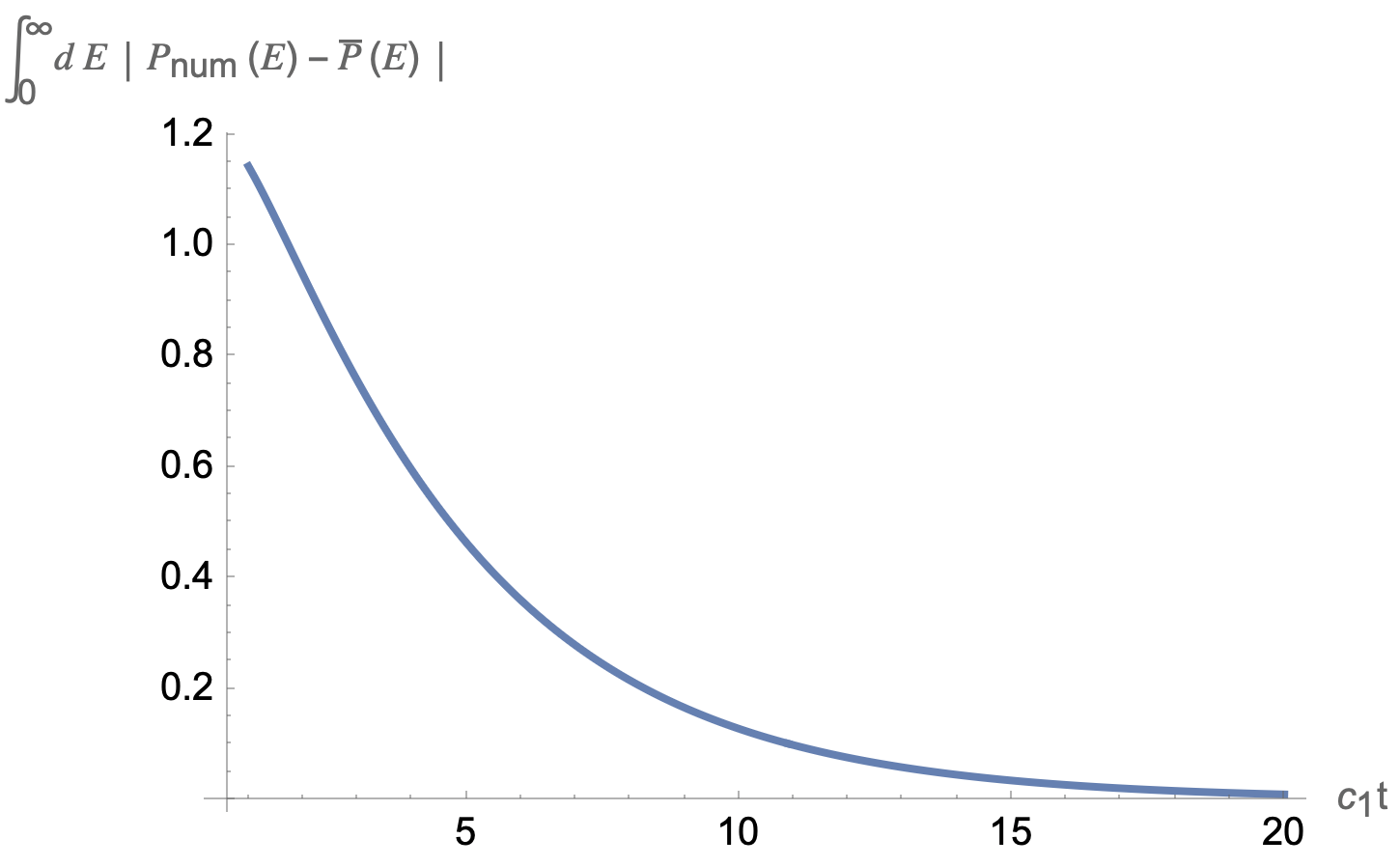}
     \captionof{figure}{The difference $\Delta P(t) \equiv \int_{0}^{\infty}dE | P_{\text{num}}(E) - \bar P(E) | $ between the numerical and attractor solutions plotted in Figure \ref{a1Pplots} as a function of time, again in units where $\Eb = 1$. The timescale required for $\Delta P<10^{-2}$ is $c_{1} t \sim 20$.}\la{a1diff}
 	 \end{center}

An interesting property of the attractor solution is that $E \bar P(E,t)$ only depends on the product $z \equiv \mathcal{E} \tau$ rather than $E$ and $t$ individually. In fact, we can see that this will be the case at long times from studying the equation \nref{PQeqn}. 

 For this, let us write the equation more generally, in a manner that is also applicable when we consider the emission of particles with spin. In general, we will find that the probability evolution equation in the quantum regime $E \ll E_{\text{brk}}$ has the form
	\begin{align}\la{Peqngen}
		\frac{1}{c_{a}}	\frac{dP(E,t)}{dt} &= -\int_{0}^{E}dE' \sqrt{E'}(E-E')^{a}P(E,t) + \int_{E}^{\infty}dE' \sqrt{E}(E'-E)^{a}P(E',t)\\
	&=-\frac{\sqrt{\pi}}{2}\frac{\Gamma(a+1)}{\Gamma(a+5/2)}E^{\frac{3}{2}+a}P(E,t) + \int_{E}^{\infty}dE' \sqrt{E}(E'-E)^{a}P(E',t)\nonumber 
	\end{align}
where $a$ is a positive integer and $c_{a}$ is a constant with mass dimensions $[c_{a}] = -(a + 1/2)$. The cases of scalar, photon, and di-photon emission correspond to $a = 1, 3,$ and $7$ respectively. Defining the variables
\begin{align}
	\mathcal{E} = E^{\frac{3}{2} + a}~~~~~~~~ \tau = \frac{\sqrt{\pi}}{2}\frac{\Gamma(a+1)}{\Gamma(a+5/2)}c_{a}t ~~~~~~~~ z = \mathcal{E} \tau
\end{align}
and performing the field redefinition
\begin{align}
	\tilde P \equiv  E P
\end{align}
\nref{Peqngen} can be written as
\begin{align}
	\frac{1}{\mathcal{E}}\frac{d \tilde P(\mathcal{E},\tau)}{d\tau} = -\tilde P(\mathcal{E}, \tau) + \frac{4}{\sqrt{\pi}(3+2a)}\frac{\Gamma(a+\frac{5}{2})}{\Gamma(a+1)}\int_{\mathcal{E}}^{\infty}\frac{d\mathcal{E}'}{\mathcal{E}'}\lb \lp \frac{\mathcal{E}'}{\mathcal{E}}\rp^{\frac{2}{3+2a}}-1\rb^{a}\tilde P(\mathcal{E}',\tau)
\end{align}
We now change variables from $(\mathcal{E}, \tau)$ to $(z,y \equiv  \mathcal{E}_{0} \tau)$ where $E_{0}$ is the energy scale of the initial distribution. 
\begin{align}
	\frac{\partial \tilde P}{\partial z} + \frac{y }{z}\frac{\partial \tilde P}{\partial y} = - \tilde P + \frac{4}{\sqrt{\pi}(3+2a)}\frac{\Gamma(a+\frac{5}{2})}{\Gamma(a+1)}\int_{z}^{\infty}\frac{dz'}{z'}\lb \lp \frac{z'}{z}\rp^{\frac{2}{3+2a}}-1\rb^{a}\tilde P(z',y)
\end{align}
As before we assume that, in the domain where $\tilde P$ has support, $z$ is order one. We will check this assumption in our final solutions and find that it is valid. In the long time limit where $y \to \infty$, the term proportional to $\partial_{y}\tilde P$ gets an infinite coefficient while the other terms remain order one. To leading order in $y$  we have the equation
\begin{align}
	\frac{y }{z}\frac{\partial \tilde P}{\partial y } \simeq 0 ~~~~~~~~~~ y \to \infty
\end{align}
This means that at long times, the $y$ dependence of $\tilde P$ drops out, and $\tilde P$  becomes a function of $z$ only. The $z$ dependence is governed by the remaining terms  which were subleading in $y$:
\begin{align}\la{zeqn}
	\frac{\partial \tilde P}{\partial z}  = - \tilde P + \frac{4}{\sqrt{\pi}(3+2a)}\frac{\Gamma(a+\frac{5}{2})}{\Gamma(a+1)}\int_{z}^{\infty}\frac{dz'}{z'}\lb \lp \frac{z'}{z}\rp^{\frac{2}{3+2a}}-1\rb^{a}\tilde P(z')
\end{align}

The photon and di-photon emission processes we will consider next involve solving \nref{Peqngen} for larger values of $a$, which complicates the analysis. The long-time equation \nref{zeqn} is somewhat simpler since it depends on one variable rather than two. We observed that in the $a = 1$ case, \nref{zeqn} admits a solution which corresponds to an attractor of the original equation \nref{Peqngen}. Our strategy for the higher spin cases will therefore be to search for a similar solution. That is, we will start by solving  \nref{zeqn} and then check that the solution is the attractor of the original equation by comparing to the numerical solution of \nref{Peqngen}.

The remainder of this section is devoted to sketching how we solve \nref{zeqn}. 


\subsubsection{Solving for the attractor}

The strategy for solving \nref{zeqn} will be as follows. We take derivatives with respect to $z$ until it is reduced to an ordinary differential equation,
\begin{align}\la{genpow}
	(z^{\frac{3}{3+2a}}\partial_{z})(z^{\frac{5+2a}{3+2a}}\partial_{z})^{a}(\partial_{z}+1)\tilde P(z) = \frac{2}{\sqrt{\pi}}\lp \frac{-2}{3+2a}\rp ^{a+1}\Gamma(a+5/2)\tilde P(z)
\end{align}
This can be transformed into a generalized hypergeometric equation, to which the solutions are known. However, 
most solutions of \nref{genpow} will not be solutions of \nref{zeqn}. In particular, when plugged into \nref{zeqn}, some solutions will diverge under the integral. The correct solution will be one which decays sufficiently quickly at large $z$ so as to be finite under the integral. This is our boundary condition at $z = \infty$.

For the boundary condition at $z = 0$, we note that the normalization condition $\int_{0}^{\infty} dE P(E) = 1$ corresponds to
\begin{align}\la{normcond}
	\int_{0}^{\infty}\frac{dz}{z}\tilde P(z) = \frac{1}{2}(3+2a)
\end{align}
which implies that
\begin{align} \la{bdzero}
	\tilde P(z = 0) = 0
\end{align}
However, it turns out that finiteness of the integral at $z = \infty$ is the only boundary condition we need to impose; once we do so \nref{bdzero} will be satisfied automatically. 

This is because \nref{zeqn} behaves essentially like a first order differential equation. In arriving at \nref{genpow} we took $a+1$ derivatives, which means that $a+1$ of the integration constants in the solution are a consequence of those derivatives. There is only one integration constant remaining, which corresponds to the single boundary condition we get to fix.

We will now illustrate the strategy we have just outlined by solving \nref{zeqn} with $a = 1$ to find \nref{Pbar}. Photon and di-photon emission correspond to $a = 3$ and $a = 7$, respectively. The derivation of the solution in those cases is conceptually analogous to the $a = 1$ case but more complicated in detail. So, in the following sections \ref{PEa3} and \ref{PEa7} we will relegate the details of those derivations to Appendices and discuss only the final solution, with the understanding that the concepts are the same as in the scalar case. 

When $a = 1$, the differential equation for $\tilde P = E P$ is
\begin{align}
	z^{2} \tilde P ''' + z^{2}\tilde P '' + \frac{7}{5}z\tilde P '' + \frac{7}{5}z \tilde P ' - \frac{3}{5}\tilde P = 0
\end{align}
which can be rewritten as
\begin{align}
	\vartheta\lp \vartheta - 1\rp \lp \vartheta-\frac{3}{5} \rp \tilde P +z (\vartheta+1) \lp \vartheta - \frac{3}{5} \rp \tilde P = 0, ~~~~~~~~\vartheta = z \partial_{z}
\end{align}
Performing a field redefinition $\tilde P = z^{\alpha}p$ shifts all the numbers in parenthesis by $+\alpha$. We see that we can remove the terms with no derivatives by setting $\alpha = 3/5$:
\begin{align}\la{peqns}
	\lp \vartheta +\frac{3}{5}\rp \lp \vartheta-\frac{2}{5} \rp \vartheta p +z \lp \vartheta+\frac{8}{5}\rp  \vartheta p = 0
\end{align}
We now have an equation for $\partial_{z}p$, effectively reducing the order of the differential equation by one. We will solve for $\partial_{z}p$ and integrate at the end to find our solution, choosing the integration constant so that the solution expanded at large $z$ has no constant part. 
\nref{peqns} admits two linearly independent solutions for $\partial_{z}p$,
\begin{align}
	\partial_{z}p = c_{1} z^{-3/5}e^{-z} + c_{2} z^{-8/5}\lp 1 + z e^{-z} \int_{z}^{\infty}\frac{dt}{t}e^{-t} \rp 
\end{align}
The second solution, integrated once with respect to $z$, is not finite under the integral of \nref{zeqn}, so we discard it. We recognize the integral of the first solution as the incomplete gamma function, $p = \Gamma\lp \frac{2}{5},z\rp$,
so we have
\begin{align}
	\tilde P \propto  z^{3/5}\Gamma \lp \frac{2}{5},z\rp 
\end{align}
Finally, we fix the normalization using \nref{normcond}.
The final answer, properly normalized, is then
\begin{align}\la{tildePa1}
	\tilde P = \frac{3}{2}z^{3/5}\Gamma \lp \frac{2}{5},z\rp 
\end{align}
which is what we found previously in \nref{Pbar}. A plot of \nref{tildePa1} can be found in Appendix \ref{Ptildeplots}, confirming that $\tilde P$ has support in the range $0 < z \lesssim 4$. 

\section{Evolution under photon emission from $j = 1/2$ black hole}\la{PEa3}

As argued in \cite{Brown:2024ajk}, when the black hole is at energies $E \lesssim E_{\text{brk}}$, it loses energy either by emission of photons or emission of photon pairs in angular momentum singlet states. We briefly summarize why this is the case and refer the reader to \cite{Brown:2024ajk} for more details. Schwarzian corrections imply that at energies below $E_{\text{brk}}$, the black hole can only occupy angular momentum states $j = 0$ or $j = 1/2$. For the $j = 1/2$ black hole, single photon emission is allowed and decay into the $\ell = 1$ photon mode is the dominant channel.  However in a $j = 0$ state, angular momentum conservation forbids the black hole from emitting a single photon, so instead it emits entangled pairs of photons with no angular momentum. Positrons are emitted stochastically on timescales which are exponentially long in the initial charge. With each positron emission, since the positron has spin one half, the black hole alternates from fermionic to bosonic or vice versa, and likewise from one type of emission to the other.

This motivates us to study the time evolution of $P(E,t)$ under these two decay channels. We first consider single photon emission from the fermionic black hole. 

When the black hole has angular momentum $j = 1/2$, the density of states is not \nref{rho} but rather \cite{Heydeman:2020hhw}
\begin{align}\la{rhoj}
	\rho_{j = \frac{1}{2}}(E,Q) = \frac{e^{S_0}}{ \pi^{2}E_{\text{brk}}}\sinh(2 \pi \sqrt{2} \sqrt{\frac{E-E_{0}^{j=1/2}}{E_{\text{brk}}}})\Theta(E-E_{0}^{j=1/2})
\end{align}
In particular, the energy spectrum is shifted up. Rather than starting at $M = Q$, the spectrum starts at $M = Q+E_{0}^{j=1/2} = \frac{3}{8}E_{\text{brk}}$. We will use
\begin{align}
	\e \equiv E - E_{0}^{j = \frac{1}{2}}
\end{align}
 to denote the energy above $E_{0}^{j=1/2}$. The decay rate of a black hole at fixed energy $E \ll \Eb $ and $j = 1/2$ due to photon emission is given in \nref{otherrates}. 
The equation we want to solve is then
\begin{align}\la{p3OGeqn}
	\frac{1}{c_{3}}\frac{d P(\e,t)}{dt} =  - \int_{0}^{\e}d \e ''  P(\e,t)\sqrt{\e''}(\e - \e '')^{3} + \int_{\e}^{\infty}d\e ' P(\e ') \sqrt{\e} (\e' - \e)^{3}  \\
\text{ where } ~~~~~~~	c_{3} =  \frac{2\sqrt{2}}{9 \pi^{2}}r_{+}^{8}E_{\text{brk}}^{9/2}~~~~~~~~~~~~~~~~~~~~~~~~~~~~~~~~~~~~~~~~~~~~~~~~~~~\nonumber
\end{align}
The solution strategy is the same as in the scalar case. We begin with \nref{genpow} for $a = 3$. For convenience we can perform a field redefinition to reduce the order of the equation we need to solve.  Among the resulting solutions, we identify the linear combination which decays sufficiently quickly at large $z$. The final answer has the form
\begin{align}\la{p3sol}
	\tilde P(z) = &v_{0} z^{1/3} + v_{1} z \, _3F_3\left(\frac{2}{3},\frac{4}{3}-\frac{i \sqrt{26}}{9},\frac{4}{3}+\frac{i \sqrt{26}}{9};\frac{11}{9},\frac{13}{9},\frac{5}{3};-z\right) \\
	+& v_{2} z^{5/9}\, _3F_3\left(\frac{2}{9},\frac{8}{9}-\frac{i \sqrt{26}}{9},\frac{8}{9}+\frac{i \sqrt{26}}{9};\frac{5}{9},\frac{7}{9},\frac{11}{9};-z\right)\nonumber \\
	+& v_{3} z^{7/9}\, _3F_3\left(\frac{4}{9},\frac{10}{9}-\frac{i \sqrt{26}}{9},\frac{10}{9}+\frac{i \sqrt{26}}{9};\frac{7}{9},\frac{11}{9},\frac{13}{9};-z\right)\nonumber
\end{align}
where $z$ and $\tilde P$ are related to the original variables by
\begin{align}\la{zdef}
	\tilde \varepsilon = \varepsilon^{\frac{9}{2}}~~~~~~~~ \tau = \frac{32}{315}c_{3}t ~~~~~~~~ z = \tilde \varepsilon \tau~~~~~~~~\tilde P \equiv  \varepsilon P
\end{align}
The $v_{i}$ are constant, order one coefficients.  Expressions for the coefficients and details of the derivation can be found in Appendix \ref{a3details}. 

We can now verify that \nref{p3sol} is the solution to \nref{p3OGeqn} at long times by comparing it to the numerical solution. In Figure \ref{a3Pplots} we plot the numerical solution to \nref{p3OGeqn} beginning from an initial delta function distribution at $\varepsilon_{0} = E_{\text{brk}}$. Visually we see the distribution approaching the attractor solution over time.

The relevant timescale is about two orders of magnitude larger than in the case of scalar emission. This is because the distribution is most naturally a function of $z$, and has support when $z$ is order one. For the scalar, $z = c_{1}t E^{5/2}$, so for $E$ to decrease by a factor of 10, $t$ must increase by a factor of $10^{5/2} \approx 3 \times 10^{2}$. For photon emission, $z = c_{3} t E^{9/2}$, so $t$ must increase by a factor of $10^{9/2} \approx 3 \times 10^{4}$. Because the distribution is evolving more slowly, it also takes longer for the numerical solution to converge to the attractor. 
As shown in Figure \ref{a3diff}, it takes $\Eb^{9/2}c_{3}t\sim 2 \times 10^{3}$ for the difference $\Delta P(t)$ to be less than $1 \%$. 

 \begin{center}
  \includegraphics[width=0.9\textwidth]{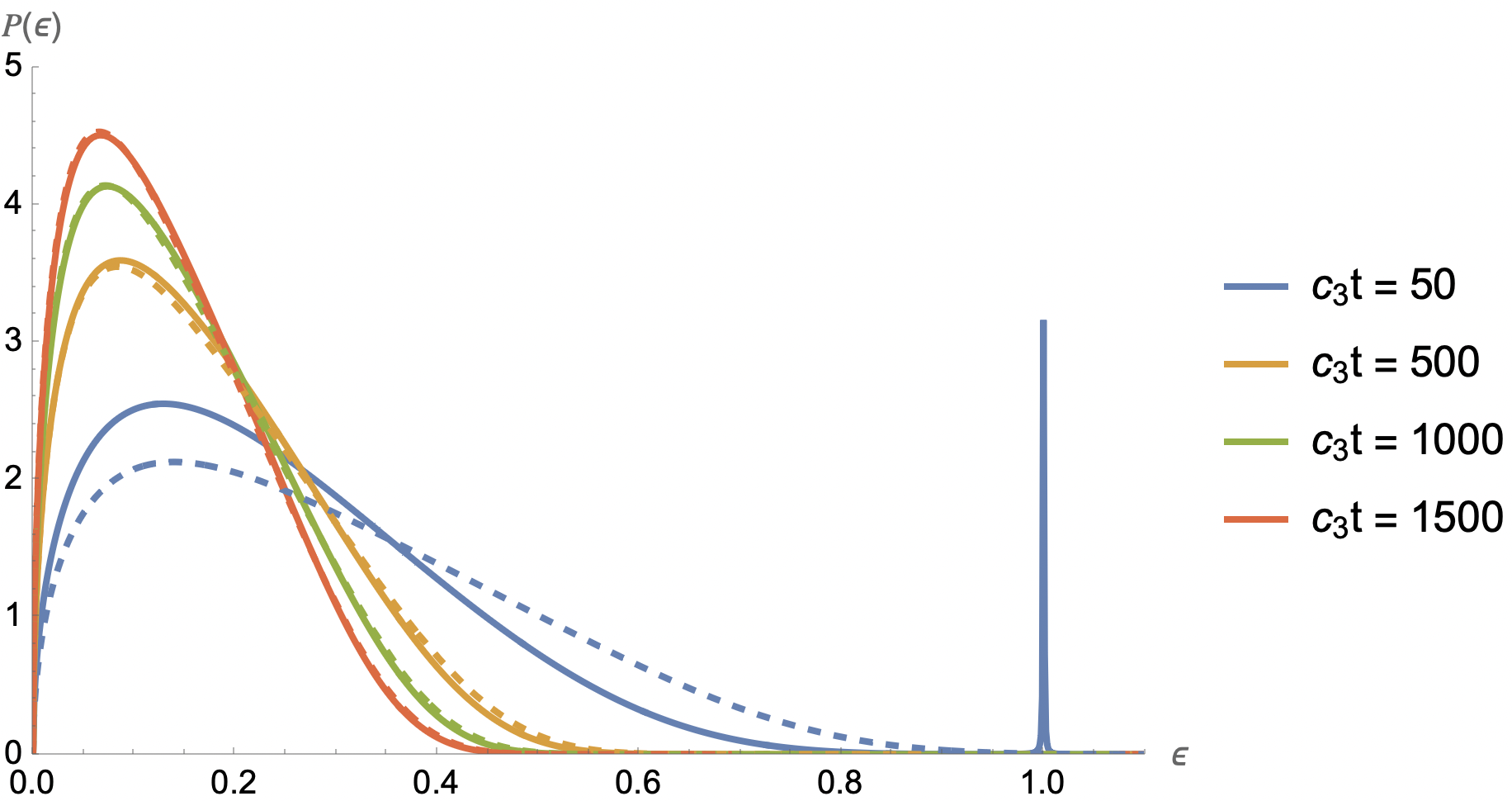}
    \captionof{figure}{Time evolution of $P(\varepsilon,t)$ for a $j = 1/2$ black hole undergoing single photon emission below the breakdown scale. Plots are in units where $E_{\text{brk}} = 1$. Solid lines: The numerical solution to \nref{p3OGeqn} starting from a delta function distribution at $\varepsilon_{0} = E_{\text{brk}}$. Dashed lines: The attractor solution $\bar{P}(\varepsilon)$ \nref{p3sol}.}\la{a3Pplots}
 	 	\end{center}

 	 	\begin{center}
  \includegraphics[width=0.7\textwidth]{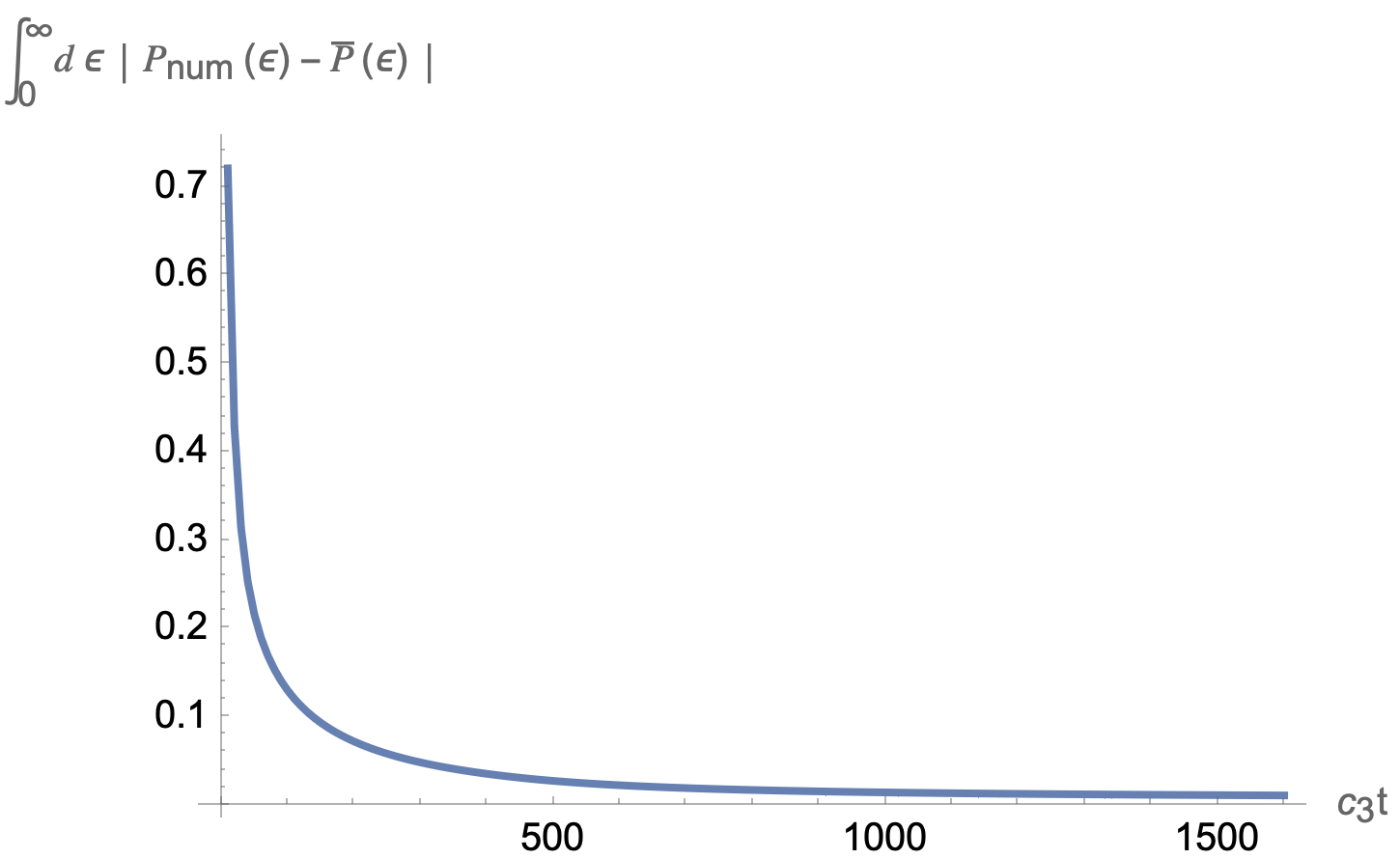}\la{a3diff}
     \captionof{figure}{The difference $\Delta P(t) \equiv \int_{0}^{\infty}d\varepsilon | P_{\text{num}}(\varepsilon) - \bar P(\varepsilon) | $ between the numerical and attractor solutions plotted in Figure \ref{a3Pplots} as a function of time in units where $\Eb = 1$. The timescale required for $\Delta P<10^{-2}$ is $c_{3} t \sim 1.6 \times 10^{3}$.}
 	 \end{center}

\section{Evolution under di-photon emission from $j = 0$ black hole}\la{PEa7}

The decay rate for a black hole at fixed energy $E \ll \Eb$ due to di-photon emission is given in \nref{otherrates}.
The equation we need to solve in this case is then
\begin{align}\la{p7OGeqn}
	\frac{1}{c_{7}}\frac{dP(E,t)}{dt} =  -\int_{0}^{E} dE'P(E,t)\sqrt{E'}(E-E')^{7}  + \int_{E}^{\infty}dE'' P(E'',t)\sqrt{E}(E''-E)^{7}\\
\text{ where } ~~~~~~	c_{7} \equiv (8.2 \times 10^{-4}) \times \frac{640\sqrt{2}}{189 \pi^{4}}E_{\text{brk}}^{17/2}r_{+}^{16}~~~~~~~~~~~~~~~~~~~~~~~~~~~~~~~~~~~~~~~~~~~~\nonumber
	\end{align}
Di-photon emission is a second order process in perturbation theory. The numerical prefactor $8.2 \times 10^{-4}$ in the transition rate, here written as part of $c_{7}$, comes from evaluating a JT gravity four-point function \cite{Brown:2024ajk}.

We find that the solution of \nref{genpow} for $a = 7$ which satisfies our boundary conditions is
	\begin{equation}\la{p7sol}
	\begin{split}
		\tilde P(z) = v_{0} z^{\alpha_0} + \sum_{j=1}^{6}v_{j}z^{\alpha_0+j \alpha_1}\, _7F_7\lp \begin{matrix}
		\alpha_{1}+(j-1)\alpha_1,&\alpha_2+(j-1)\alpha_1,&...&\alpha_{7}+(j-1)\alpha_1\\
		\alpha_0+j\alpha_1,& \alpha_0+(j+1)\alpha_1,& ...*...&\alpha_0+(j+8)\alpha_1
	\end{matrix}; -z\rp \\
	+v_{7}z \, _7F_7\lp \begin{matrix}
		7 \alpha_{1},&\alpha_2+6\alpha_1,&...&\alpha_{7}+6\alpha_1\\
		\alpha_0+8\alpha_1,& \alpha_0+9\alpha_1,& ....&\alpha_0+14\alpha_1
	\end{matrix}; -z\rp ~~~~~~~~~~~~~ \\
\end{split}
	\end{equation}
	where
	\begin{equation}
  \begin{split}
  	\alpha_{0} &= \frac{3}{17}\\
	\alpha_{1} &= \frac{2}{17}\\
	\alpha_{2} &= \frac{12}{17} - \frac{1}{17}\sqrt{49 + r_{1}}\\
	\alpha_{3} &= \frac{12}{17} + \frac{1}{17}\sqrt{49 + r_{1}}
	\end{split} \quad \quad \quad 
	\begin{split}
	\alpha_{4} &= \frac{12}{17} - \frac{1}{17}\sqrt{49 + r_{2}}\\
	\alpha_{5} &= \frac{12}{17} + \frac{1}{17}\sqrt{49 + r_{2}}\\
	\alpha_{6} &= \frac{12}{17} - \frac{1}{17}\sqrt{49 + r_{3}}\\
	\alpha_{7} &= \frac{12}{17} + \frac{1}{17}\sqrt{49 + r_{3}}
  \end{split}
\end{equation}
and $r_{1},r_{2},r_{3}$ denote the 3 roots of the polynomial
	\begin{align}\la{rs}
		675675+12345x+163x^{2}+x^{3}
	\end{align}
The  $*$  in \nref{p7sol} indicates that the $\alpha_0+7\alpha_1 = 1$ term is always omitted from the sequence of parameters of the hypergeometric function.
The coefficients $v_{i}$ can be found in Appendix \ref{a7details}. 

We again confirm that \nref{p7sol} is the correct long-time solution of \nref{p7OGeqn} by comparison to numerics. Here $z = c_{7}t E^{17/2}$, so the timescale for the energy to decrease by a factor of 10 is now $10^{17/2} \approx 3 \times 10^{8}$, which is reflected in Figure \nref{a7Pplots}, as well as  in the time required for the numerical and attractor solutions to converge. We find that $\Eb^{17/2} c_{7} t \sim 10^{10}$ before $\Delta P(t)<10^{-2}$.

	 \begin{center}
    \includegraphics[width=0.9\textwidth]{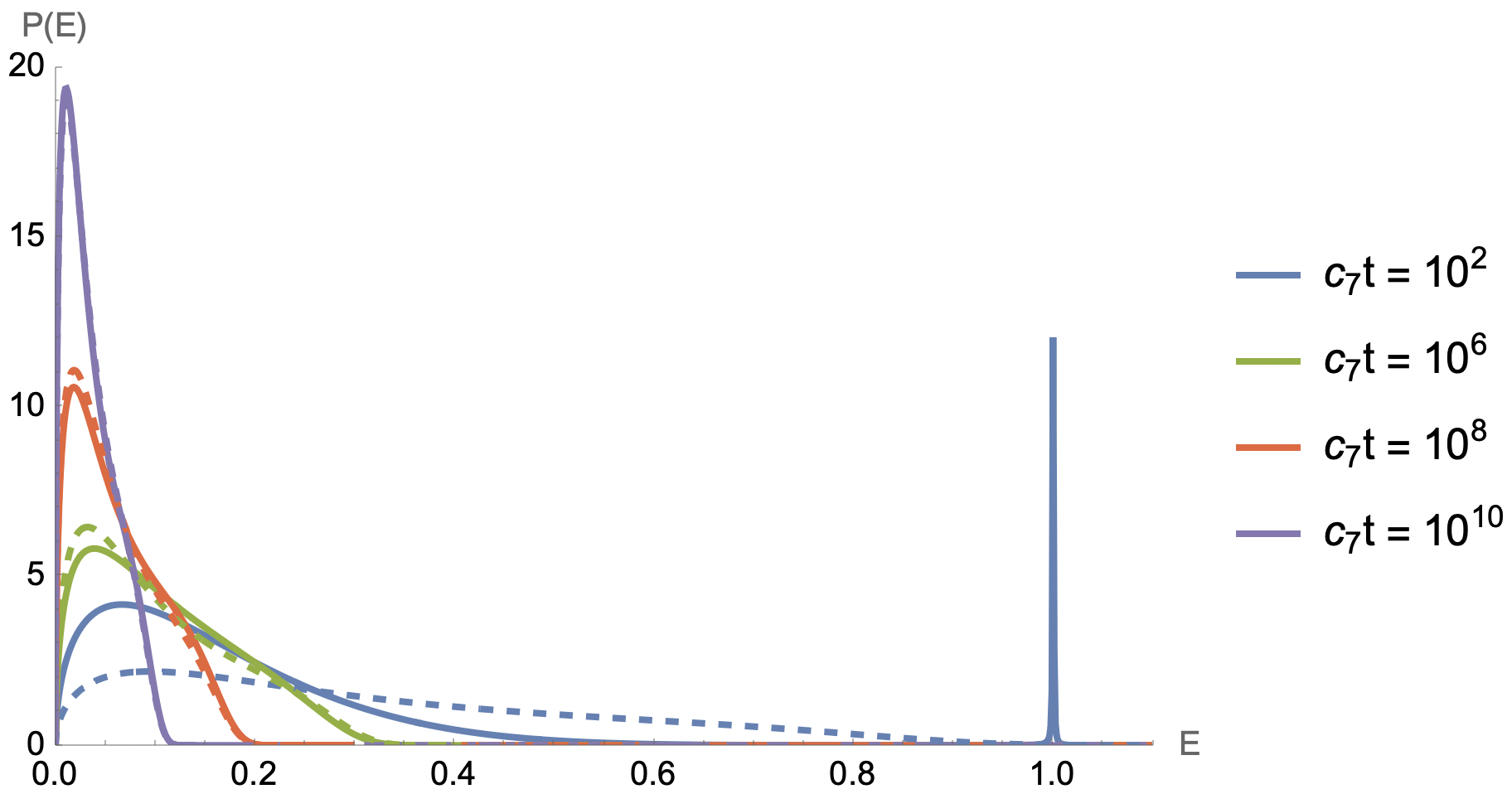}
     \captionof{figure}{Time evolution of $P(E,t)$ for a $j = 0$ black hole undergoing di-photon emission below the breakdown scale. Plots are in units where $E_{\text{brk}} = 1$. Solid lines: The numerical solution to \nref{p7OGeqn} starting from a delta function distribution at $E_{0} = E_{\text{brk}}$. Dashed lines: The attractor solution $\bar{P}(E)$ \nref{p7sol}.}\la{a7Pplots}
 \end{center}
 
 	 \begin{center}    \includegraphics[width=0.8\textwidth]{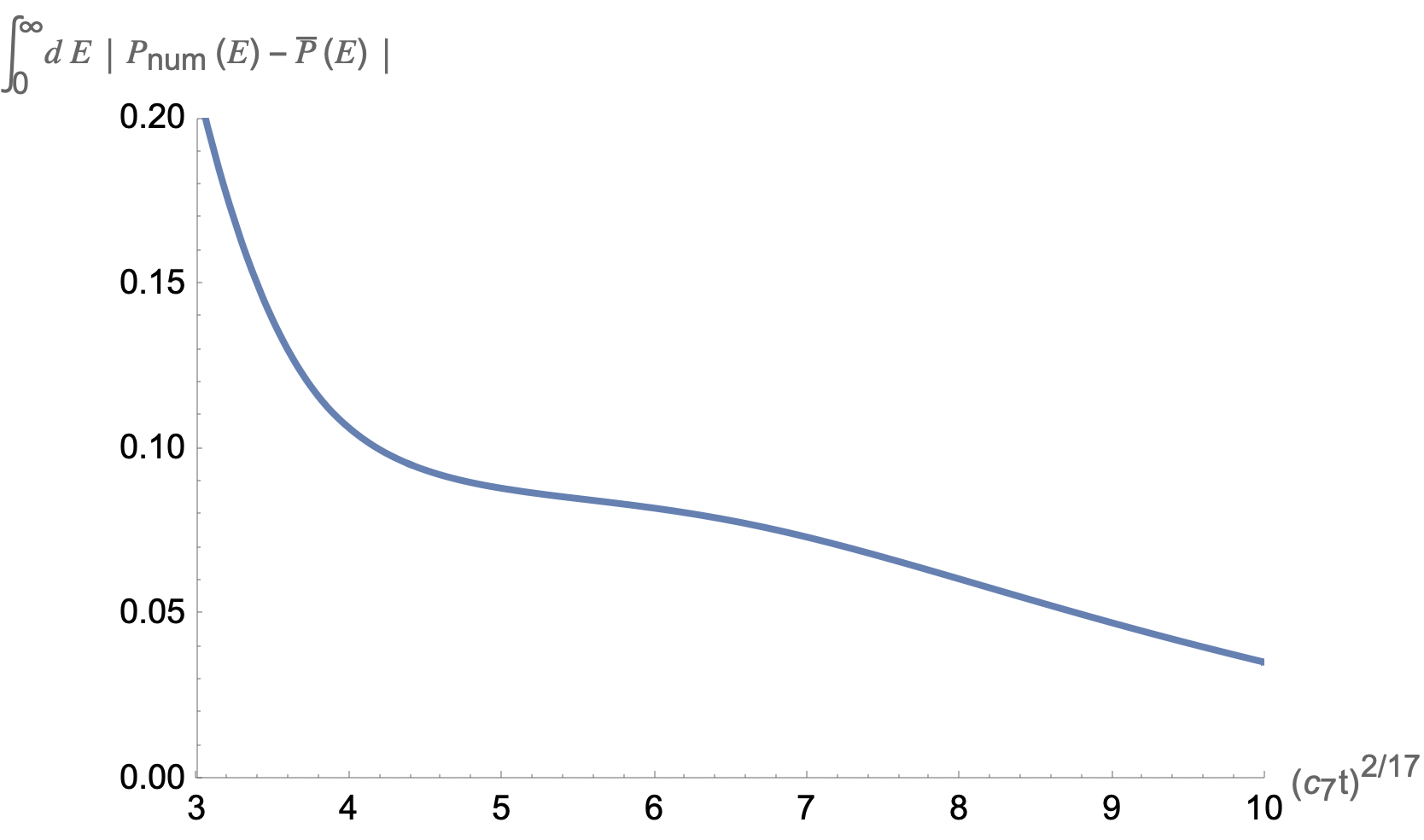} \la{a7diff}
     \captionof{figure}{ The difference $\Delta P(t) \equiv \int_{0}^{\infty}d\varepsilon | P_{\text{num}}(\varepsilon) - \bar P(\varepsilon) | $ between the numerical and attractor solutions shown in figure \ref{a7Pplots}. Here we plot $\Delta P$ as a function of  $(c_{7}t)^{2/17}$, which is the natural time variable measuring order one changes in $\langle E \rangle$.  The timescale required for $\Delta P(t) < 10^{-2}$ is $c_{7} t \sim 10^{10}$ (not shown).}
 	  \end{center}

\section{Corrected spectrum and emission rates}\la{Spec}

In this section we present results for the expected energy $\langle E(t)\rangle$ of the black hole, the energy flux $\langle dE/dt\rangle $, and the Hawking emission spectra in the various attractor states. 

First, we note that in all three cases, the long time solution exhibits a kind of scaling symmetry which fixes the time dependence of $\langle E(t) \rangle$, and likewise the dependence of $\langle dE/dt\rangle$ on $\langle E \rangle$. Since $E \bar P$ depends only on $z$, it is invariant under the rescaling
\begin{align}
	t \to \eta t ~~~~~~~~~~~~ E \to \eta^{-\frac{2}{3+2a}} E
\end{align}
At long times, the expected energy is an integral over this function with respect to $E$, so the time dependence can be scaled out by a simple change of variables:
\begin{align}
	\langle E(t) \rangle &= \int_{0}^{\infty}dEE \bar P(E,t)\\
	&\propto \frac{1}{t^{\frac{2}{3+2a}}}\int_{0}^{\infty} \frac{dz}{z^{\frac{2a+1}{2a+3}}}f(z)\nonumber 
\end{align}
where $f(z)$ is some function of $z$.

Therefore in the attractor solution, $\langle E(t)\rangle$ has a power law time dependence given by $t^{-2/5}$, $t^{-2/9}$, and $t^{-2/17}$ when the black hole is undergoing scalar emission, photon emission, and di-photon emission, respectively. These are the same powers found by computing the energy flux $dE/dt$ in a microcanonical state and integrating with respect to time \cite{Brown:2024ajk}. We now compute the overall coefficients.

  \subsection{Scalar emission}
   For neutral particle emission, the expected energy and energy flux at long times is
\begin{align}
   	\langle E(t)\rangle   
   	&=\frac{3}{5} \lp \frac{15}{4} \rp^{2/5} \Gamma\left(\frac{7}{5}\right)\frac{1}{(c_{1}  t)^{2/5}}\la{expEs}\\
   \langle \frac{d E }{dt} \rangle 
   &=-\frac{8 }{27}\sqrt{\frac{5}{3}} \, \Gamma
   \left(\frac{7}{5}\right)^{-5/2}c_{1}
   \langle E \rangle^{7/2}
     	\text{for } ~~~\langle E \rangle \ll \Eb, ~~\Eb^{5/2}c_{1}t \gg 1 \la{dEdts}
\end{align}


  We can compare this to the microcanonical result. The  energy flux from a black hole in a state of fixed energy $E_{i}$ is \cite{Brown:2024ajk}
\begin{align}\la{dEdtmicros}
	\frac{d  E }{dt}\bigg|_{E_{i}} 
	&=-\frac{16 }{105 }c_{1}E_{i}^{\frac{7}{2}} ~~~~~~ E_{i} \ll E_{\text{brk}}
\end{align}
The corresponding microcanonical probability distribution  is $\tilde P(z) = \frac{5}{2}z \delta(z - \frac{7}{10})$, or written in the original variables, $P(E,t) = \delta(E - E(t)) $ where $E(t) = \lp \frac{21}{8} \rp^{2/5}(c_{1}t)^{-2/5}$. 

Comparing \nref{dEdtmicros} to \nref{dEdts} at the expected energy $\langle E \rangle = E_{i}$, we find that the energy flux in the attractor state is larger  than the one in the microcanonical ensemble by a factor of $\sim 3.4$.
%
%
This difference is also reflected in the Hawking spectrum. The microcanonical emission spectrum for neutral scalar particles in the low energy limit is \cite{Brown:2024ajk}
\begin{align}\la{dNdtdwmic}
	\frac{dN}{dt d\omega } \bigg|_{E_{i}}
	&=c_{1}\omega \sqrt{E_{i}-\omega}\Theta(E_{i}-\omega) ~~~~~~ E_{i} \ll E_{\text{brk}}
		\end{align}		
		We find the particle spectrum in the attractor state \nref{Pbar} by integrating it against the microcanonical result,
\begin{align}
\langle \frac{dN}{dtd\omega}\rangle &=	c_{1}\int_{0}^{\infty}dE \bar P(E)\omega \sqrt{E-\omega}\Theta(E-\omega) \nonumber \\
&= \frac{1}{50} \lp \frac{3^{4}}{2^{4}\, 5}\rp^{1/10}
   \omega^3( c_{1}^8 t^3)^{1/5}
   G_{5,7}^{7,0}\left(\frac{4}{225} c_{1}^2 t^2
   \omega ^5 \bigg|
\begin{array}{c}
   -\frac{1}{10},\frac{1}{10},\frac{3}{
   10},\frac{1}{2},1 \\
   -\frac{2}{5},-\frac{1}{5},0,0,\frac{
   1}{5},\frac{1}{5},\frac{2}{5} \\
\end{array}
\right)
\end{align}
where $G$ denotes the Meijer G function. We can also express it as a function of the expected energy \nref{expEs},
%
\begin{align}
	\langle \frac{dN}{dtd\omega}\rangle 
&=\frac{9 \sqrt{3} \Gamma
   \left(\frac{7}{5}\right)^{3/2}}{500\ 2^{3/5}
   } c_{1} \omega^3 \langle E \rangle ^{-3/2}
   G_{5,7}^{7,0}\left(\frac{243
   \Gamma
   \left(\frac{7}{5}\right)^5}{12500
  }\frac{\omega^{5}}{\langle E \rangle^{5}}\bigg|
\begin{array}{c}
   -\frac{1}{10},\frac{1}{10},\frac{3}{
   10},\frac{1}{2},1 \\
   -\frac{2}{5},-\frac{1}{5},0,0,\frac{
   1}{5},\frac{1}{5},\frac{2}{5} \\
\end{array}
\right) \la{dNdtdwPbar}
\end{align}
At frequencies small compared to $\langle E \rangle$, \nref{dNdtdwPbar} reduces to
\begin{align}
	\langle \frac{dN}{dtd\omega}\rangle = 2^{7/10} \sqrt{\frac{3 \pi }{5 \left(5+\sqrt{5}\right)}} \Gamma \left(\frac{11}{10}\right)\Gamma \left(\frac{7}{5}\right)^{-3/2}c_{1}\langle E \rangle^{1/2} \omega  ~~~~~~~~~~ \omega \ll \langle E \rangle \ll E_{\text{brk}}
\end{align}
which has the same frequency dependence as \nref{dNdtdwmic} in this limit. In Figure \ref{dNdtdwplt1} we plot the particle flux per unit frequency in a microcanonical state of energy $E_{i} = 10^{-2}\Eb$ and in the attractor state at expected energy $\langle E \rangle = 10^{-2} \Eb$. We also plot the energy flux per unit frequency, which simply differs from the particle flux per unit frequency by a factor of $\omega$. 

The microcanonical spectrum ends at $\omega = E_{i}$ because the black hole cannot emit more energy than it has. Of course, we do not see this cutoff in the $\bar P$ spectrum because the black hole has some probability of being in many different states. 

	 \begin{center}
      \includegraphics[width=0.8\textwidth]{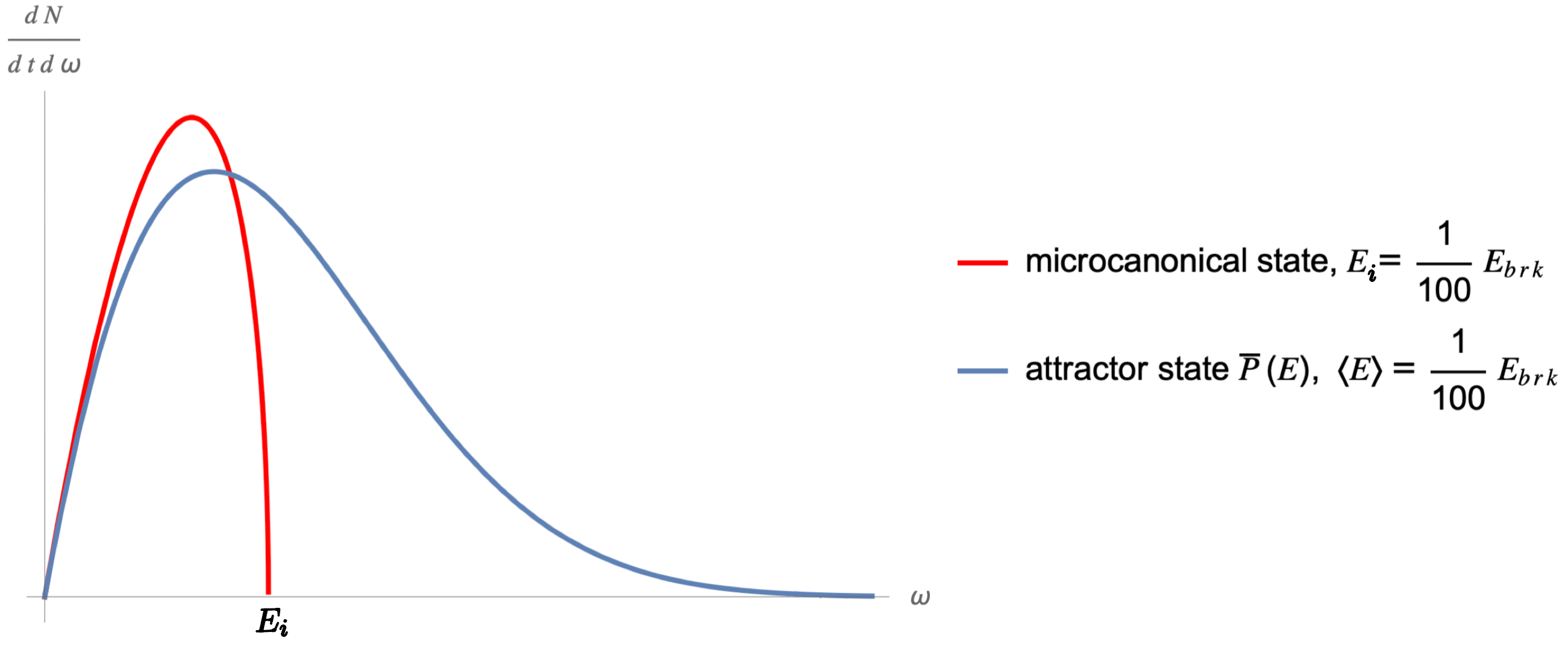}
      
        \textbf{(a)}
    
  \includegraphics[width=0.8\textwidth]{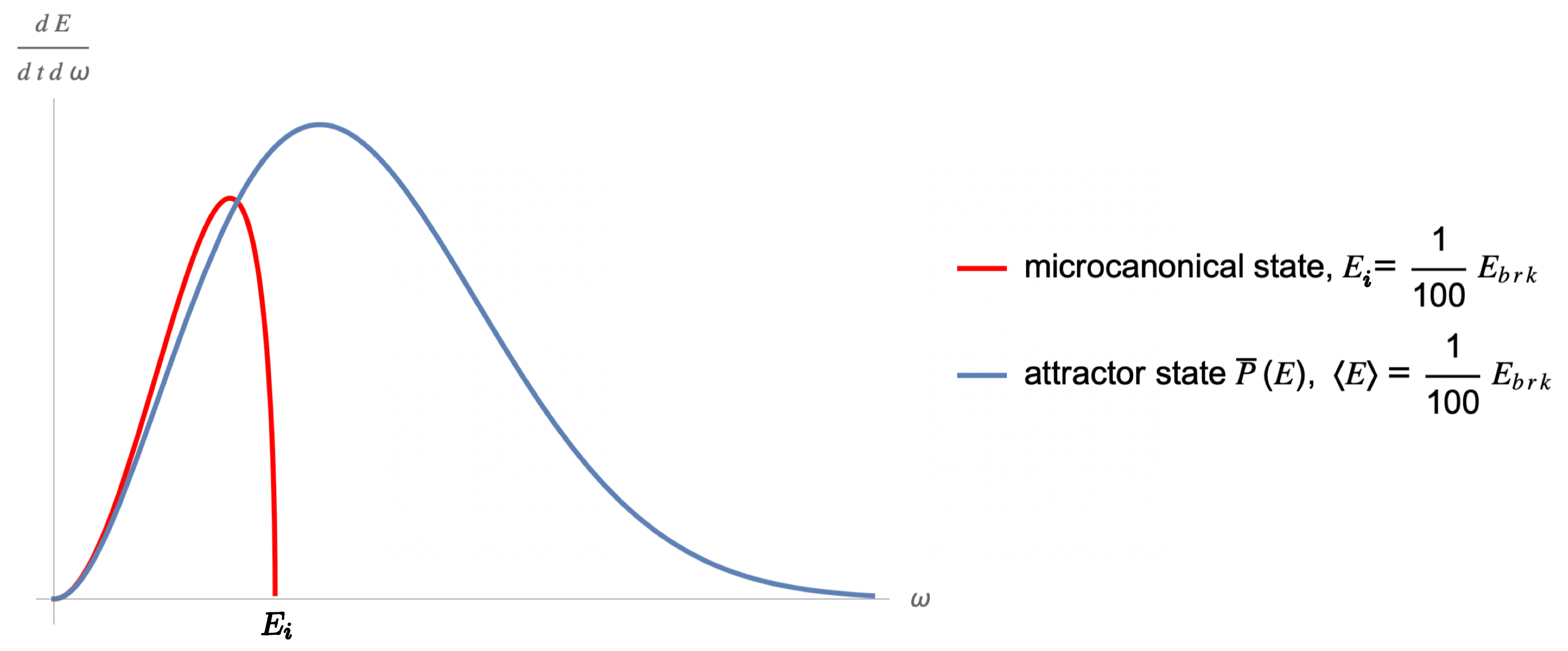}
  
   	 \textbf{(b)}
   	 
   	    \captionof{figure}{Hawking radiation into the $\ell = 0$ massless scalar mode at energy $\langle E \rangle = \frac{1}{100} \Eb$. We plot the spectrum in the distribution \nref{Pbar} which the black hole would occupy at long times (blue), compared to that in a microcanonical fixed energy state (red). a) Expected particle flux per unit frequency. b) Expected energy flux per unit frequency.  }\la{dNdtdwplt1}

 	 \end{center}

\subsection{Photon emission from $j = 1/2$ black hole}
The expected energy and energy flux of $\ell = 1$ photons from a fermionic black hole below the breakdown scale is
\begin{align}
		\langle \varepsilon(t) \rangle 
&=\frac{21\ 3^{4/9} 35^{2/9} \pi  \cos \left(\frac{\pi }{18}\right) \Gamma \left(\frac{11}{9}\right) \Gamma \left(\frac{22}{9}\right)}{2^{1/9} \, 26 \left(1+2 \cosh
   \left(\frac{2 \sqrt{26} \pi }{9}\right)\right) \Gamma \left(\frac{8}{9}\pm \frac{i \sqrt{26}}{9}\right) \Gamma \left(\frac{4}{3}\pm \frac{i \sqrt{26}}{9}\right) }\frac{1}{(c_{3}t)^{2/9}}\\
\langle \frac{d \varepsilon }{dt} \rangle 
&=-\frac{1827904 \sqrt{\frac{13}{21}}}{551353635} \left(\frac{\left(1+2 \cosh \left(\frac{2 \sqrt{26} \pi
   }{9}\right)\right) \Gamma \left(\frac{8}{9}\pm \frac{i \sqrt{26}}{9}\right) \Gamma \left(\frac{4}{3}\pm \frac{i \sqrt{26}}{9}\right)}{\cos \left(\frac{\pi }{18}\right) \Gamma \left(\frac{11}{9}\right) \Gamma \left(\frac{22}{9}\right)}\right)^{9/2}c_{3}\langle \varepsilon \rangle^{11/2}\la{dedta3} \\
&\text{ for } ~~\langle E \rangle  \ll E_{\text{brk}}, ~~\varepsilon \ll \langle E \rangle,~~~ \Eb^{9/2}c_{3}t\gg 1\nonumber 
\end{align}
Here we have evaluated the observables in the attractor state \nref{p3sol}. We can again compare this to a black hole in a microcanonical state with the same expected energy, $E_{i} = \langle E \rangle$. In the microcanonical ensemble, the energy flux is \cite{Brown:2024ajk}
\begin{align}\la{dedtmic3}
	\frac{d \varepsilon}{dt}\bigg|_{\varepsilon_{i}} 
	&= -\frac{256}{3465}c_{3}\varepsilon_{i}^{11/2}
\end{align}
In this case, \nref{dedta3} is larger than \nref{dedtmic3} by a factor of $\sim 14.3$. 

The microcanonical emission spectrum is \cite{Brown:2024ajk}
\begin{align}\la{dNdtdwmica3}
	\frac{dN}{dt d\omega}\bigg|_{\varepsilon_{i}} 
		&=c_{3}\, \omega^{3}\sqrt{\varepsilon_{i}-\omega}~\Theta(\varepsilon_{i}-\omega)~~~~~~~~\varepsilon_{i} \ll E_{i},~~ E_{i} \ll \Eb
\end{align}
The functional form of \nref{p3sol} is sufficiently complicated that we will not attempt to write down an analytic expression for the emission spectrum, but instead just plot it numerically. The results are shown in Figure \ref{a3plots} for states with $\langle \varepsilon \rangle = 10^{-2}\Eb$.

	 \begin{center}
  \includegraphics[width=0.8\textwidth]{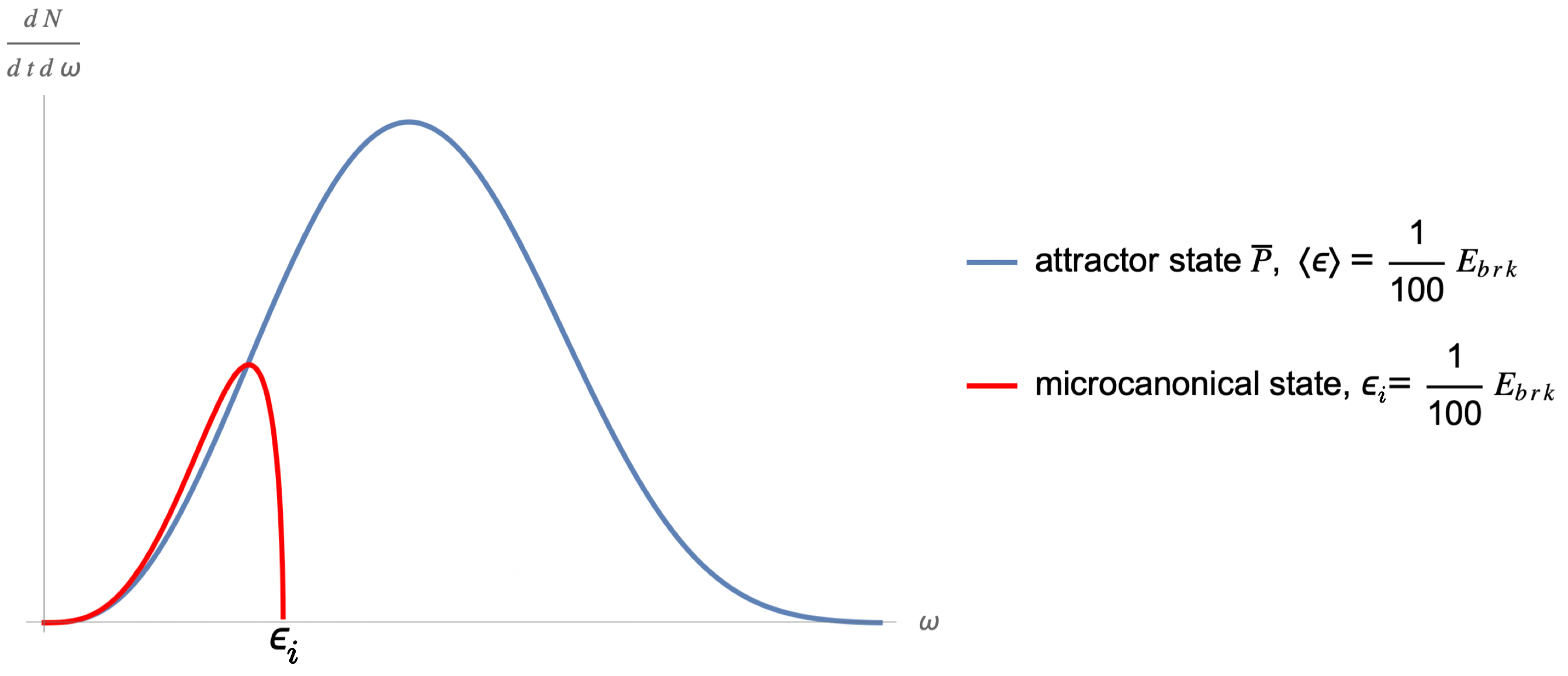}
  
  \textbf{(a)}
  
  \includegraphics[width=0.8\textwidth]{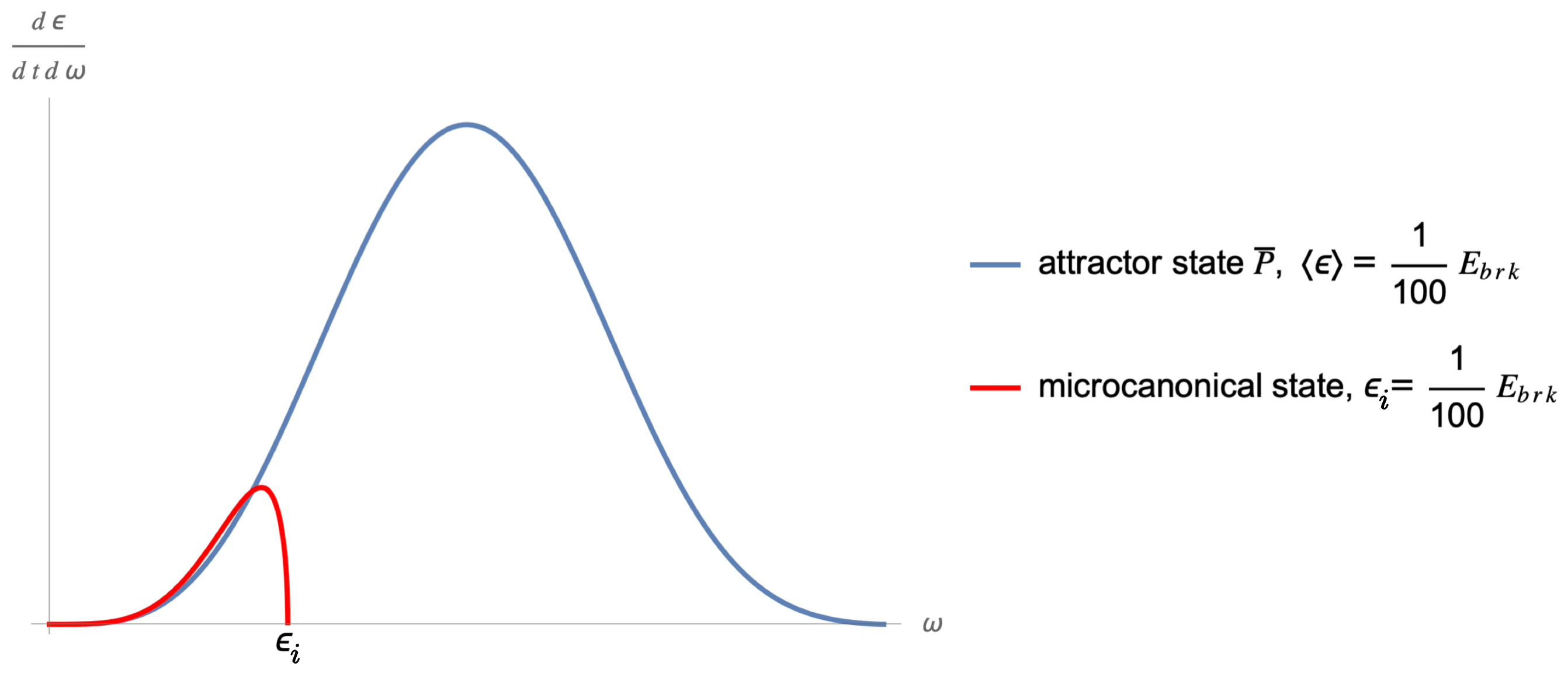}
  
  \textbf{(b)}
  
     \captionof{figure}{Hawking radiation into $\ell =1$ photons from a black hole with $j = 1/2$ and energy   $\langle E \rangle = \frac{1}{100} \Eb$. In blue we plot the spectrum in the distribution \nref{p3sol}, which is the state the black hole would occupy at long times. The microcanonical spectrum is shown in red. a) Expected particle flux per unit frequency. b) Expected energy flux per unit frequency.  }\la{a3plots}

 	 \end{center}

 	 We see that the emission rates per unit frequency peak at a frequency larger than the expected energy of the black hole state.
 	  To understand this, we note that the microcanonical emission rate \nref{dNdtdwmica3} is maximized at the frequency $\omega_{\text{max}} = \frac{6 \varepsilon_{i}}{7}$, where it has the value $\frac{dN}{dtd\omega}\big|_{\varepsilon_{i}} \propto \varepsilon_{i}^{7/2}$. So, the magnitude of the contribution to the total emission rate increases steeply with the energy of the microcanonical state. This means that when we integrate over microcanonical states with the probability distribution \nref{p3sol}, larger energy states contribute more at each $\omega$ than smaller energy states. We will see the same effect present in the next section.

\subsection{Di-photon emission from $j = 0$ black hole}

Finally, in the case of di-photon emission, we find that the expected energy and energy flux in the attractor state \nref{p7sol} are 
\begin{align}
	\langle E (t) \rangle &= 0.541649 \times \frac{1}{(c_{7}t)^{2/17}}\\
	\langle \frac{d  E }{dt} \rangle &= -21.5765 \times c_{7} \langle E \rangle^{19/2}
	\la{dEdta7}
~~~~~~~\text{ for }~\langle E \rangle \ll \Eb, ~~~~ \Eb^{17/2}c_{7}t \gg 1
\end{align}
where we have given numerical approximations for the constant prefactors.

For comparison, the microcanonical emission rate is \cite{Brown:2024ajk}
\begin{align}\la{a7micro}
	\frac{dE}{dt}\bigg|_{E_{i}} &= -\frac{65536}{2078505}c_{7}E_{i}^{19/2}~~~~~ E_{i} \ll E_{\text{brk}}
\end{align}
\nref{dEdta7} is larger than \nref{a7micro} by a factor of $\sim 684$.

The microcanonical emission spectrum is \cite{Brown:2024ajk}
\begin{align}
	\frac{dN}{dt d\omega}\bigg|_{E_{i}}
	&=c_{7}\, \omega^{7}  \sqrt{E_{i}-\omega}~\Theta(E_{i}-\omega)~~~~~~ E_{i} \ll \Eb
\end{align}
In Figure \ref{p7plots} we have numerically evaluated the emission spectra in the attractor state \nref{p7sol}, again at energy $\langle E \rangle = 10^{-2} \Eb$. The microcanonical spectrum is shown for comparison. We zoom in on this portion of the plot to make it visible relative to the spectrum of the attractor state. 

	 \begin{center}\la{p7plots}
    \includegraphics[width=0.9\textwidth]{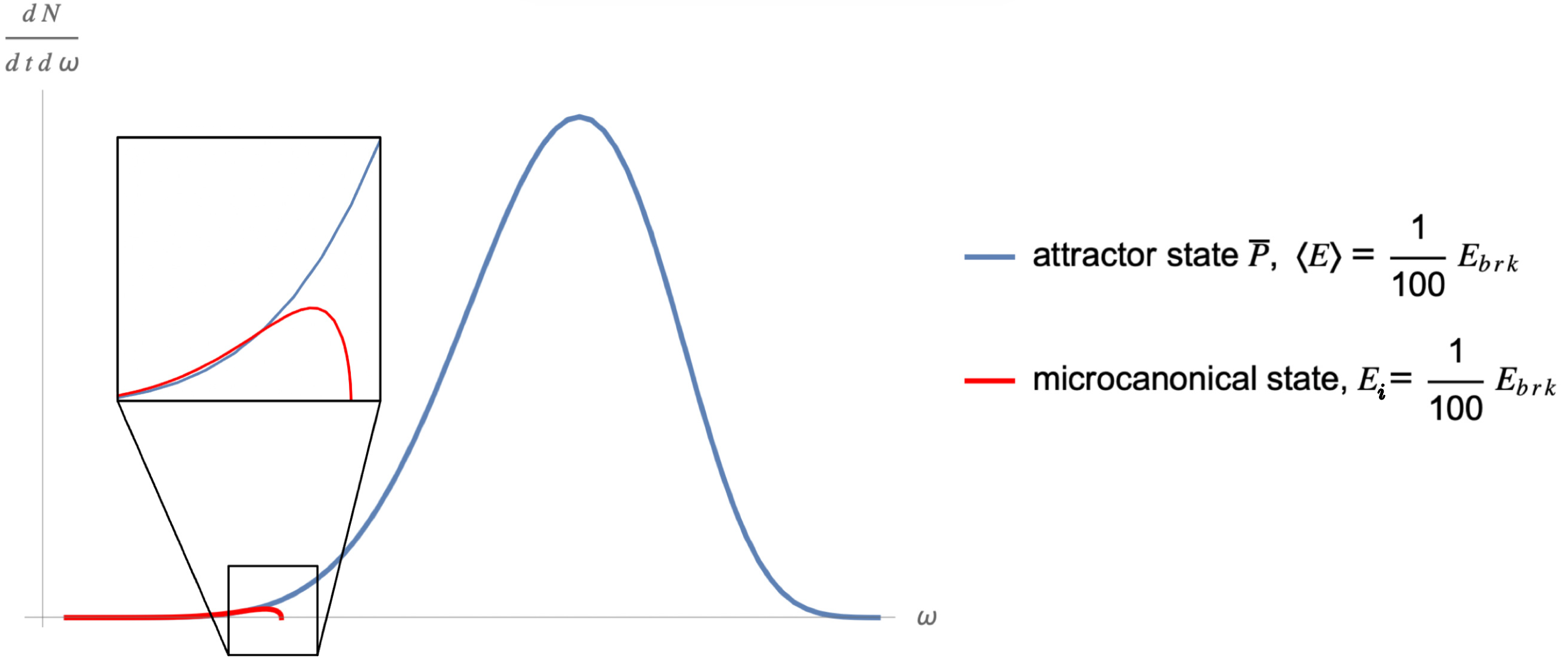}
    
    \textbf{(a)}
    
  \includegraphics[width=0.9\textwidth]{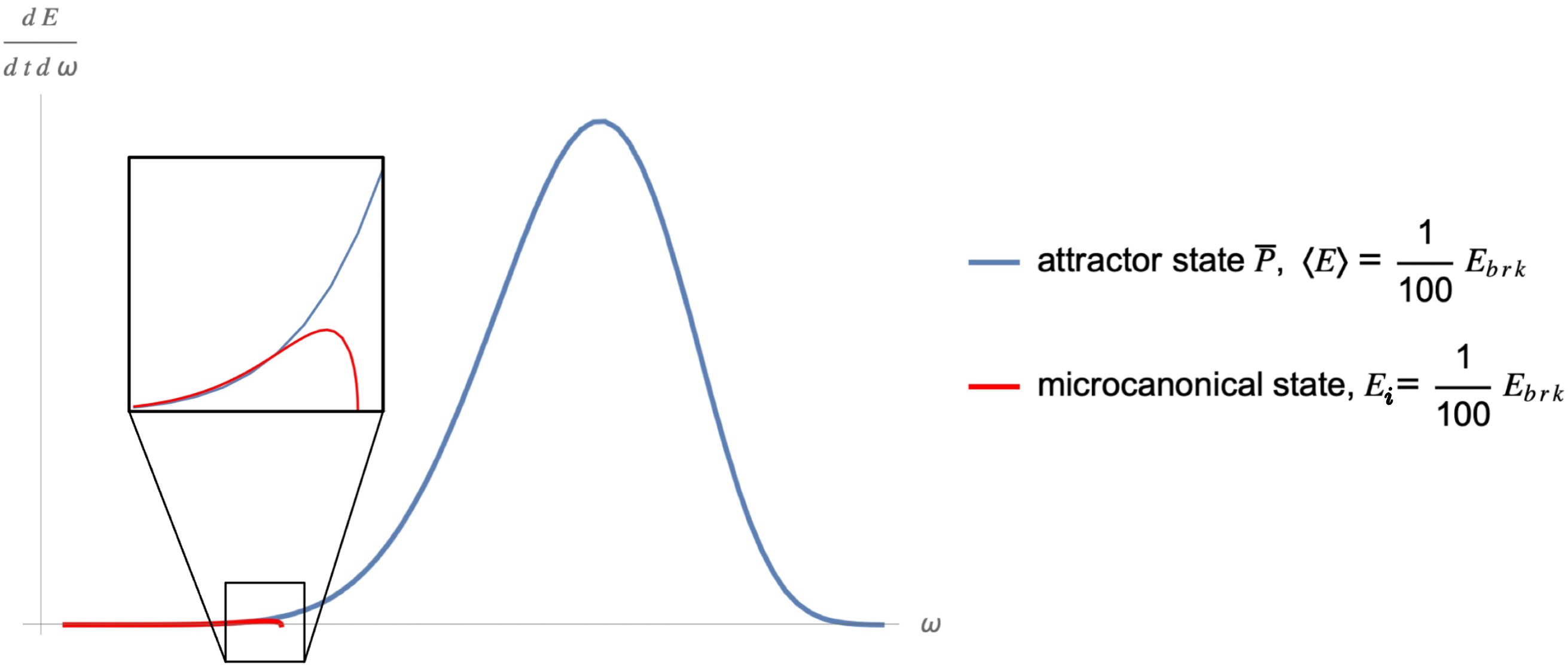}
  
  \textbf{(b)}
  
   \captionof{figure}{Hawking radiation into entangled singlet states from a black hole with $j = 0$ and energy  $\langle E \rangle = \frac{1}{100} \Eb$. In blue we plot the spectrum in the distribution \nref{p7sol}. The microcanonical spectrum is shown in red for comparison. a) Expected particle flux per unit frequency. b) Expected energy flux per unit frequency.  }

 	 \end{center}

\section{Absorption cross section}\la{ACS}

In this section we comment on Schwarzian corrections to another observable, the absorption cross section. A similar discussion can be found in \cite{Emparan:2025sao}, which appeared while this paper was in preparation.

We will focus on the simplest case, that of a neutral massless scalar field. Suppose $\phi$ begins in a coherent state at frequency $\omega$ and with expected particle number $N_{\omega}$. The expected number of particles absorbed per unit time is the difference between the ingoing and outgoing flux. By a derivation similar to the one in section \ref{ETemission}, this can be expressed in terms of $\langle OO \rangle$ as
\begin{align}\la{absform}
	\Phi_{\text{in}} - \Phi_{\text{out}} = -\frac{g^{2}}{2 \omega}G(-\omega) + \frac{g^{2}N_{\omega}}{2 \omega} \lp G(\omega) - G(-\omega) \rp ~~~~~~~~ g = 2 r_{+}
\end{align}
where we have used $G(\omega)$ to denote the Wightman function,
\begin{align}
G(\omega) \equiv \int_{-\infty}^{\infty}dt e^{i \omega t}\langle O(t) O(0)\rangle
\end{align}
The derivation of \nref{absform} can be found in Appendix \ref{sigmaabs}.

The first term of \nref{absform} describes the effect of spontaneous emission, as can be seen from the expression for the vacuum emission probability in section \ref{ETemission}. The effect of spontaneous emission is small compared to the other two terms when $N_{\omega} \gg 1$. 

The second and third terms describe absorption and stimulated emission, respectively. \nref{absform} involves more than just the second term, which we might call ``pure absorption,'' because we assume that a classical measurement would not distinguish between a mode which had been reflected and one which had been emitted.

As was the case for the emission rate, the leading Schwarzian corrections to the absorption cross section are determined simply by expanding $G(\omega)$ in the limit where the black hole energy or temperature is much below the breakdown scale. To given an example, let us consider a black hole in a thermal state and a scalar field with large quantum number $N_{\omega} \gg 1$. In this limit Hawking radiation can be neglected. To go from \nref{absform} to the absorption cross section, we divide by the ingoing particle flux. 
\begin{align}
	\sigma_{\text{abs}}(\omega) = \frac{2 r_{+}^{2}}{ \omega} \lp G(\omega) - G(-\omega)\rp ~~~~~~~~~~ N_{\omega} \gg 1
\end{align}
The KMS condition implies that $G(-\omega) = e^{-\omega \beta}G(\omega)$, so \nref{absform} becomes
\begin{align}\la{abssc}
	\sigma_{\text{abs}}(\omega) = \frac{2 r_{+}^{2}}{ \omega}(1-e^{- \beta \omega })G(\omega) ~~~~~~~~~~~ N_{\omega} \gg 1
\end{align}
As a sanity check, we can see whether this formula reproduces the area in the semiclassical limit. In this limit, $G(\omega)$ is fixed by conformal symmetry to be the $\Delta = 1$ correlator \nref{confFT},
\begin{align}
	G(\omega) = \frac{2 \pi \omega}{1-e^{-\beta \omega}}~~~~~~~~ 1/\beta \gg \Eb,~~ \omega \ll E
\end{align}
which plugged into \nref{abssc} immediately gives $4 \pi r_{+}^{2}$. 

In the canonical ensemble, the two point function is given by
\begin{align}\la{Gcan}
	G(\omega)_{\text{can}}  &= 2 \pi Z(\beta)^{-1} \int_{0}^{\infty} dE ~e^{-\beta E } \rho(E) \rho(E+\omega) |O_{E, E+\omega}|^{2} 
\end{align}
where the partition function is obtained from the density of states \nref{rho} by a Laplace transform. Approximating the integral in the limit $\omega \ll \beta^{-1} \ll \Eb$, we find
\begin{align}
	\sigma_{\text{abs}}^{\text{can}}(\omega) \simeq 4 r_{+}^{2}\sqrt{\frac{2 \beta \Eb}{\pi}} \gg 4 \pi r_{+}^{2} ~~~~~~~~~~~~~~~\omega \ll 1/\beta \ll \Eb
\end{align}
This is parametrically larger than the semiclassical prediction.

A similar calculation using the microcanonical correlator \nref{Gwmic} shows that the absorption cross section in a fixed energy state is also enhanced by Schwarzian corrections:
\begin{align}
	\sigma_{\text{abs}}^{\text{mic}}(\omega) \simeq 4 r_{+}^{2} \sqrt{\frac{\Eb}{2 E_{i}}}  \gg 4 \pi r_{+}^{2} ~~~~~~~~~~~~~~~~~~~~ \omega \ll E_{i} \ll \Eb
\end{align}

As we have stressed, observables such as the absorption cross section depend on the energy probability distribution that they are evaluated in. As seen from \nref{absform}, it also depends on the occupation number of the incoming wave. We could wonder whether there is a universal signature of the Schwarzian which is independent of these considerations. One such limit is
\begin{align}\la{lim}
	\beta^{-1}  \sim \langle E \rangle \ll \omega \ll \Eb 
\end{align}
In this regime the absorption cross section is not sensitive to the details of the energy probability distribution. For example, in this limit the canonical and microcanonical correlators \nref{Gcan} and \nref{Gwmic} coincide. Moreover, \nref{absform} simplifies because the terms containing $G(-\omega)$  drop out, and the absorption cross section is independent of $N_{\omega}$. For a near-extremal black hole in the Schwarzian regime, these are the first type of Schwarzian corrections we would notice as we lower the frequency of the incoming wave. 

In the limit \nref{lim} we find that the scalar absorption cross section is enhanced, this time by a frequency-dependent factor,
\begin{align}\la{absnew}
\sigma_{\text{abs}}(\omega) \simeq A_{H} \frac{1}{\pi}\sqrt{\frac{E_{\text{brk}}}{2 \omega}}~~~~~~~~~~~~~\beta^{-1} \sim E \ll \omega \ll E_{\text{brk}}
\end{align}
We can wonder how quickly the absorption of scalar modes with frequency \nref{lim} would bring the black hole energy above $\Eb$, at which point Schwarzian corrections would be negligible and \nref{absnew} would no longer be correct. In 4d, the absorption cross section is related to the s-wave absorption probability by $\mathcal{P} \propto \omega^{2}\sigma$, so the increase in energy due to absorption of a single mode scales as $\delta E \propto \omega^{3} \sigma$. If $\phi$ has occupation number $N_{\omega}$, then we need $N_{\omega} \delta E \ll \Eb$ which, using \nref{absnew} for $\sigma$, implies
\begin{align}
	N_{\omega} \ll \sqrt{\frac{\Eb}{\omega}}\frac{1}{(r_{+}\omega)^{2}}
	\end{align}
This is not a strict bound, given that both factors on the righthand side are much greater than one in the limit under consideration.

	 \begin{center}
     \includegraphics[width=0.8\textwidth]{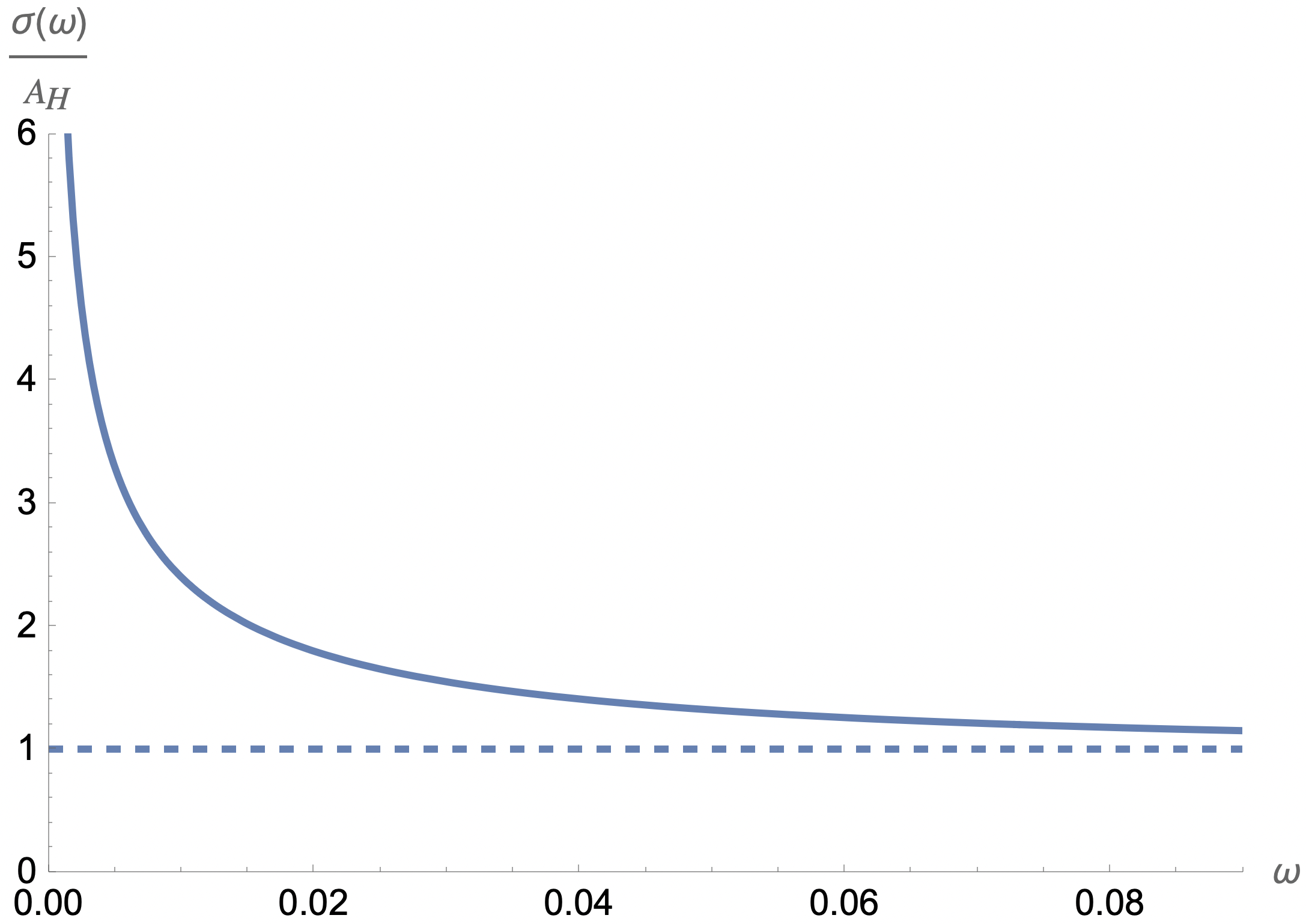}
     \captionof{figure}{Absorption cross section in the limit $\beta^{-1} \sim E \ll \omega \ll E_{\text{brk}}$, plotted in units where $E_{\text{brk}}  = 1$. Note that this plot stops being correct when $\omega$ becomes of order $E$. }
 	 \end{center}

\section{Conclusion}

In this paper we studied the energy probability density $P(E,t)$ of the black hole as it evolves towards extremality in the deep quantum gravity regime.
We found that, below the breakdown scale $\Eb$ where the Schwarzian becomes strongly coupled, the state of the black hole evolves toward a non-thermal, universal long-time distribution. Fortunately,  in order to correctly predict the behavior of the black hole state at late times, we do not need to know the details of its entire evolution history above $\Eb$, as one might have thought. We also found that the attractor solutions effectively depend only on one combination of energy and time which fixes  the powers of time and energy in $\langle E\rangle$ and $\langle dE/dt\rangle$. The Hawking fluxes calculated in the attractor state can be much larger than those in a microcanonical state with the same expected energy. In the case of di-photon emission, it is enhanced by a factor of $\sim 700$. The corrected emission spectra also have a markedly different functional form. In the case of scalar emission, the Hawking spectrum in the attractor state is given by a Meijer-G function. We also discussed a kind of ``universal'' Schwarzian correction to the scalar absorption cross section which would be independent of the particular energy probability distribution of the black hole.

This work concerned the evolving state of near-extremal charged black holes, but we could ask a similar question about near-extremal Kerr-Newman black holes. The Schwarzian-corrected evaporation rate of those black holes due to scalar emission was recently studied in \cite{Maulik:2025hax}. For the emission of particles with angular momentum, we expect that the  black hole state will evolve much differently due to the presence of instabilities in superradiant modes. We could also consider the state of near-BPS black holes in $\mathcal{N} = 2$ supergravity. In this case, there is a gap $E_{\text{gap}}$ in the spectrum between the degenerate ground states and the first excited state. So, we would expect to see the population of excited states decrease exponentially with time, with a decay rate proportional to $E_{\text{gap}}^{-1}$.

It would also be interesting to consider possible phenomenological implications  for the lifetime of near-extremal charged primordial black holes. Primordial black holes are one possible explanation for dark matter \cite{Hawking:1971ei, Carr:1974nx, Carr:2016drx, Sasaki:2018dmp}, which in some models are near-extremal and charged under a $U(1)$ gauge field in the dark sector \cite{Bai:2019zcd}. Here we have studied how Schwarzian effects would decrease the Hawking evaporation rate of such objects. These calculations may be relevant for determining the parameter space of allowed masses in models of dark matter consisting of near-extremal charged black holes.

Schwarzian effects would also change the decoherence rates of quantum systems in the exterior of near-extremal black holes. For example, the decoherence rate of quantum superpositions outside a Schwarzchild and a Kerr black hole were calculated in \cite{Danielson:2022tdw, Gralla:2023oya}. The decoherence can be understood as arising due to the black hole absorbing quanta of the fields sourced by the superposition. If the superposition sources a scalar field, we expect that the decoherence rate would be enhanced by Schwarzian effects, per the discussion in  section \ref{ACS}. To understand how the Schwarzian affects the decoherence of charged or massive particles, we would need to calculate the cross section for electromagnetic and gravitational fields, respectively. In this case, angular momentum conservation will play a role, since the absorption of a single photon or graviton by a near-extremal RN black hole increases the energy by an amount of order $\Eb$ \cite{Brown:2024ajk}. By the same logic, the ``dynamical Love number'' of near-extremal black holes, which are directly related to their absorption cross section, would be modified by Schwarzian effects \cite{Perry:2024vwz}. We leave these calculations to future work. 
\vspace{3mm}

\textbf{Acknowledgements}

We would like to thank Andreas Blommaert, Roberto Emparan, Luca Iliesiu, Guanda Lin, Geoff Penington, Olivier Simon, Mykhaylo Usatyuk, Chen Yang, and Zihan Zhou for useful discussions and/or helpful communication. We especially thank Juan Maldacena for invaluable advising and comments on the draft.

\appendix

\section{Full solution for $P(E,t)$ under scalar emission}\la{PQsol}

Here we derive the solution to the energy probability equation \nref{PQeqn}, restated here:
\begin{align}\la{inteq}
	\frac{1}{c_{1}}\frac{dP(E,t)}{dt} = -\frac{4}{15}E^{5/2}P(E)+ \sqrt{E}\int_{E}^{\infty}dE'(E'-E)P(E')
\end{align}
It will be convenient to define the following variables:
\begin{align}
	\mathcal{E} \equiv E^{5/2}, ~~~~~ \tilde T \equiv \frac{2}{5}T, ~~~~~ \tau \equiv \frac{4}{15} c_{1} t, ~~~~~z \equiv \mathcal{E} \tau
\end{align}
We expand $P$ in a basis of mode functions of the form $E^{iT}F(T,z)$. That is, we make the ansatz
\begin{align}
	P(E,t) = \int_{-\infty}^{+\infty}dT A(T) E^{iT}F(T,z)
\end{align}
where $A(T)$ is some function fixed by the initial conditions. Plugging $E^{iT}F(T,z)$ into \nref{inteq}, we have
\begin{align}\la{Finteqn}
	\frac{\partial F(T,z)}{\partial z} + F(T,z) = \frac{15}{4}\int_{1}^{\infty}dv (v-1)v^{iT}F(T, v^{5/2}z)
\end{align}
By taking derivatives with respect to $z$, we convert \nref{Finteqn} into a third-order differential equation,
\begin{align}\la{Fdiffeqn}
	\frac{5}{3} z^{\frac{1}{5}-i \tilde T} \partial_{z} \lb z^{7/5}\partial_{z} \lp z^{\frac{2}{5}+i \tilde T} \lp \partial_{z}F + F \rp \rp \rb  = F
\end{align}
Not all solutions of \nref{Fdiffeqn} will be solutions of \nref{Finteqn}. In particular, we are looking for a solution which is finite under the integral in  \nref{Finteqn}. To identify solutions which are well behaved at large $z$, we make a modified Frobenius series ansatz around $z = \infty$. $z = \infty$ is an irregular singular point of \nref{Fdiffeqn}, which suggests that solutions expanded around $z = \infty $  will have the form
\begin{align}
	F(T, x) = e^{-1/x}\sum_{n = 0}^{\infty}x^{r+n}~~~~~ x \equiv 1/z
\end{align}
The indicial equation of \nref{Fdiffeqn} around $x = 0$ gives a single exponent, $r = 0$. It can be shown that the resulting series has the integral expression
\begin{align}
	F(T, z) = e^{-z}\lp -\frac{5}{3} + \int_{1}^{\infty}\frac{dy}{y}y^{-\frac{1}{5}+i \tilde T}e^{-z(y-1)} \rp 
\end{align}
For convenience, we can define a unit normalized function satisfying $\hat F (T,z = 0) = 1$. Evaluating $F$ at $z = 0$,
\begin{align}
	F(T,z = 0) = -\frac{5}{3}+ \int_{1}^{\infty}\frac{dy}{y}y^{-\frac{1}{5}+i \tilde T} = -\frac{5}{3} +\frac{1}{\frac{1}{5}-i \tilde T} = -\frac{5}{3}\frac{\lp \frac{2}{5}+i\tilde T\rp}{(-\frac{1}{5}+i\tilde T)}
\end{align}
where in performing the integral we have assumed that $\text{Re}(-\frac{1}{5}+i \tilde T)<0$. So, we arrive at the normalized mode function
\begin{align}
	\hat F(T,z) = -\frac{3}{5}\frac{(-\frac{1}{5}+i\tilde T)}{\lp \frac{2}{5}+i\tilde T\rp}e^{-z}\lp -\frac{5}{3} + \int_{1}^{\infty}\frac{dy}{y}y^{-\frac{1}{5}+i \tilde T}e^{-z(y-1)} \rp 
\end{align}
We now impose the initial conditions. Assuming that we begin at $t = 0$ with a delta function distribution at $E_{0}$,
\begin{align}
	P(E,t = 0) = \delta(E - E_{0})
\end{align}
then $A(T)$ is fixed by
\begin{align}
\delta(E - E_{0}) = \int dT A(T) E^{iT} \quad \RA \quad A(T) = \frac{1}{2\pi}E_{0}^{-1-iT}
\end{align}
The full solution is
\begin{align}
	P(E,t) =- \frac{1}{2\pi}\frac{3}{2}\frac{1}{E_{0}}\int_{\mathcal{C}}d\tilde T\lp \frac{\mathcal{E}}{\mathcal{E}_{0}}\rp ^{i \tilde T}\frac{(-\frac{1}{5}+i\tilde T)}{\lp \frac{2}{5}+i\tilde T\rp}e^{-\mathcal{E} \tau }\lp -\frac{5}{3} + \int_{1}^{\infty}\frac{dy}{y}y^{-\frac{1}{5}+i \tilde T}e^{-\mathcal{E} \tau (y-1)} \rp
\end{align}
where we impose that the integral contour $\mathcal{C}$ lies along a straight line where $\text{Re}(\frac{2}{5}+i \tilde T)<0$.

All that remains is to perform the $\tilde T$ integral. Whether we close the contour in the upper or lower half complex $\tilde T$ plane depends on whether $E > E_{0}$ or $E < E_{0}$. Since the black hole can only transition to lower energy states, we should find that $P(E>E_{0},t) = 0$ for all $t \geq 0$. 

When $E > E_{0}$, we close the contour in the upper half plane, or where $\text{Re}(i \tilde T)<0$.  We do not pick up the pole at $i \tilde T = -\frac{2}{5}$ since we have imposed that $\mathcal{C}$ lies above this point in the complex plane, and the result of the integral is zero. When $E < E_{0}$, we close the contour in the lower half plane, or where $\text{Re}(i \tilde T)>0$. Writing
\begin{align}
	\frac{(-\frac{1}{5}+i\tilde T)}{\lp \frac{2}{5}+i\tilde T\rp} = 1-\frac{3}{5}\frac{1}{\frac{2}{5}+i\tilde T}
\end{align}
we see that the fourier transform produces a delta function from the first term and a theta function from the second. The final answer, for $E<E_{0}$, is
\begin{align*}
	P(E,t) = e^{-\mathcal{E}_{0}\tau }\delta(E-E_{0}) + \frac{3}{2}\frac{1}{E} \lp e^{-\mathcal{E} \tau } - e^{-\mathcal{E}_{0}\tau } \lp \frac{\mathcal{E}}{\mathcal{E}_{0}}\rp^{\frac{3}{5}} -\frac{3}{5}\lp \mathcal{E} \tau \rp^{\frac{3}{5}}\lb \Gamma\lp -\frac{3}{5},\mathcal{E} \tau \rp  - \Gamma\lp -\frac{3}{5}, \mathcal{E}_{0} \tau \rp \rb \rp 
\end{align*}
as stated in \nref{Pfullsol}.

\section{Derivation of attractor solutions under photon and di-photon emission}
In this appendix we solve \nref{zeqn} for $a = 3$ and $a = 7$. 

\subsection{$a=3$}\la{a3details}

The differential equation \nref{genpow} with $a = 3$ is
\begin{multline}\la{Pa3}
	\vartheta\lp \vartheta-\frac{1}{3}\rp \lp \vartheta-\frac{5}{9}\rp \lp \vartheta-\frac{7}{9}\rp \lp \vartheta-1\rp \tilde P+
	z \lp \vartheta-\frac{1}{3}\rp\lp \vartheta+1\rp \lp \vartheta+\frac{1}{9}(3+i\sqrt{26})\rp \lp \vartheta+\frac{1}{9}(3-i\sqrt{26})\rp \tilde P = 0
\end{multline}
\nref{Pa3} has the form of a generalized hypergeometric equation, 
\begin{align}\la{genhye}
	\lp \vartheta (\vartheta + b_{1} -1)...(\vartheta + b_{q}-1) + z (\vartheta + a_{1}) ... (\vartheta + a_{p}) \rp w = 0
\end{align}
which is naively solved by five $_4F_4$ functions. However, $b_{1}$ and $a_{1}$ differ by an integer, reducing the order of the $_4F_4$ functions by one, and $b_{4} = 0$, so we do not have the usual set of fundamental solutions. 

We perform a field redefinition $\tilde P = z^{1/3} p$ to get an equation for $\partial_{z}p$,
\begin{align}\la{peqn3}
	\lp \vartheta+\frac{1}{3}\rp \lp \vartheta-\frac{2}{9}\rp \lp \vartheta-\frac{4}{9}\rp \lp \vartheta-\frac{2}{3}\rp \vartheta p+	z\lp \vartheta+\frac{4}{3}\rp \lp \vartheta+\frac{2}{3}+\frac{i\sqrt{26}}{9}\rp \lp \vartheta+\frac{2}{3}-\frac{i\sqrt{26}}{9}\rp \vartheta p = 0
\end{align}
We construct the solutions to this equation by making a Frobenius series ansatz around $z = 0$. We find four solutions $\{ f_{i} \}$ for $\partial_{z}p$, three  $_2F_2$ functions and one Meijer G function. The three $_2F_2$ functions are
\begin{align}
f_{1}(z) &\equiv	z^{-1/3}\, _2F_2\left(\frac{4}{3}-\frac{i \sqrt{26}}{9},\frac{4}{3}+\frac{i
   \sqrt{26}}{9};\frac{11}{9},\frac{13}{9};-z\right)\\
  f_{2}(z) &\equiv  z^{-5/9}\, _2F_2\left(\frac{10}{9}-\frac{i \sqrt{26}}{9},\frac{10}{9}+\frac{i
   \sqrt{26}}{9};\frac{7}{9},\frac{11}{9};-z\right)\\
 f_{3}(z) &\equiv  z^{-7/9} \, _2F_2\left(\frac{8}{9}-\frac{i \sqrt{26}}{9},\frac{8}{9}+\frac{i
   \sqrt{26}}{9};\frac{5}{9},\frac{7}{9};-z\right)
\end{align}
The Meijer G function has no exponentially decaying terms at large $z$, so we ignore it.

At large $z$, the generalized hypergeometric function is a sum of terms polynomial and exponential in $z$. As $z \to \infty$, the leading order terms in the asymptotic expansion of $_qF_q$ are
\begin{align}
	_qF_q \lp a_{1},...,a_{q}; b_{1},...,b_{q}; -z\rp \sim \prod_{\ell = 1}^{q} \frac{\Gamma(b_{\ell})}{\Gamma(a_{\ell})}\sum_{m = 1}^{p}\Gamma(a_{m})\frac{\prod_{\ell = 1, \ell \neq m}^{p}\Gamma(a_{\ell}-a_{m})}{\prod_{\ell = 1}^{q}\Gamma(b_{\ell}-a_{m})}z^{-a_{m}} + e^{-z}(\text{polynomial in }z)
	\end{align}

The part proportional to $e^{-z}$ will converge under the integral. The polynomial terms which scale as $z^{-a_{m}}$ will not. Above we have written only the largest order term which is polynomial in $z$. The remaining polynomial terms  go as $z^{-a_{m} - k}$ for $k \in \Z_{>0}$ and converge under the integral. So to solve the original equation, we take a linear combination of the $\{f_{i}\}$  which eliminates all of the terms which scale as $z^{-a_{m}}$. This amounts to solving a set of linear equations for the coefficients of the above three functions.

To be completely explicit, labelling the coefficients of the asymptotic expansions of the $f_{i}$ as components of a matrix $M$,
\begin{align}
	f_{i}(z \to \infty)   = M_{1i}~z^{-\frac{5}{3}-\frac{i \sqrt{26}}{9}} + M_{2i}~z^{-\frac{5}{3}+\frac{i \sqrt{26}}{9}}  + e^{-z}(\text{polynomial in z})
\end{align}
and denoting the coefficients of our desired linear combination by $v_{i}$,
\begin{align}
	\partial_{z}p = \sum_{i} v_{i}f_{i}
\end{align}
we need to solve the equation
\begin{align}
	M.v = 0
\end{align}

Once we have the $v_{i}$, we integrate with respect to $z$ to find $p$, choosing the constant of integration so that the asymptotic expansion of $p$ at large $z$ has no constant part. Finally, imposing the normalization
\begin{align}
	\frac{2}{3+2a}\int_{0}^{\infty}\frac{dz}{z}\tilde P(z) = 1
\end{align}
we find
\begin{equation}
\begin{split}\la{tildePa3}
\tilde P(z) = &v_{0} z^{1/3} + v_{1} z \, _3F_3\left(\frac{2}{3},\frac{4}{3}-\frac{i \sqrt{26}}{9},\frac{4}{3}+\frac{i \sqrt{26}}{9};\frac{11}{9},\frac{13}{9},\frac{5}{3};-z\right) \\
	+& v_{2} z^{5/9}\, _3F_3\left(\frac{2}{9},\frac{8}{9}-\frac{i \sqrt{26}}{9},\frac{8}{9}+\frac{i \sqrt{26}}{9};\frac{5}{9},\frac{7}{9},\frac{11}{9};-z\right)\\
	+& v_{3} z^{7/9}\, _3F_3\left(\frac{4}{9},\frac{10}{9}-\frac{i \sqrt{26}}{9},\frac{10}{9}+\frac{i \sqrt{26}}{9};\frac{7}{9},\frac{11}{9},\frac{13}{9};-z\right)
\end{split}		
\end{equation}
where
\begin{align*}
   v_{0} &= \frac{15 \sin \left(\frac{2 \pi }{9}\right) \sec \left(\frac{\pi }{18}\right) \Gamma \left(\frac{2}{9}\right) \Gamma \left(\frac{2}{3}\right) \Gamma \left(\frac{11}{9}\right)  \Gamma \left(\frac{2}{3}\pm \frac{i \sqrt{26}}{9}\right)}{4 \Gamma \left(\frac{5}{9}\right) \Gamma \left(\frac{14}{9}\right) \Gamma \left(\frac{1}{3}\pm \frac{i \sqrt{26}}{9}\right)}\\
      v_{1} &= -\frac{315}{32}\\
    v_{2} &= \frac{15 \sin \left(\frac{2 \pi }{9}\right) \sec \left(\frac{\pi }{18}\right) \Gamma \left(\frac{2}{9}\right)^2  \Gamma \left(\frac{8}{9}\pm \frac{i
   \sqrt{26}}{9}\right) \left(\cos \left(\frac{2 \pi }{9}\right)-\cosh \left(\frac{2 \sqrt{26} \pi }{9}\right)\right)^2}{4 \left(\sin \left(\frac{\pi }{18}\right)-\sin \left(\frac{1}{18} \left(5+4 i
   \sqrt{26}\right) \pi \right)\right) \Gamma \left(\frac{5}{9}\right) \Gamma \left(\frac{14}{9}\right) \Gamma \left(\frac{1}{3}\pm \frac{i \sqrt{26}}{9}\right) 
   \left(-\sin \left(\frac{\pi }{18}\right)+\cos \left(\frac{2}{9} \left(\pi +i \sqrt{26} \pi \right)\right)\right)}\\
    v_{3} &= \frac{135 \, \Gamma \left(\frac{2}{9}\right) \Gamma \left(\frac{10}{9}\pm \frac{i \sqrt{26}}{9}\right) \left(\cos \left(\frac{2 \pi }{9}\right)-\cosh \left(\frac{2
   \sqrt{26} \pi }{9}\right)\right)^2}{16 \left(\cos \left(\frac{2 \pi }{9}\right)-\cos \left(\frac{2}{9} \left(2+i \sqrt{26}\right) \pi \right)\right) \Gamma \left(\frac{14}{9}\right) \Gamma
   \left(\frac{1}{3}\pm \frac{i \sqrt{26}}{9}\right) \left(\cos \left(\frac{2 \pi }{9}\right)-\sin \left(\frac{1}{18} \left(\pi +4 i \sqrt{26} \pi
   \right)\right)\right)}
\end{align*}

\subsection{$a = 7$}\la{a7details}

When $a = 7$, the differential equation \nref{genpow} for  $\tilde P$ can be written as the generalized hypergeometric equation \nref{genhye} with
\begin{equation}
  \begin{split}
   p&=q=8\\
	a_{1} &= 1\\
	a_{2} &= -3/17\\
	a_{3} &= -\frac{1}{17}\lp -7-\sqrt{49+r_{1}}\rp \\
	a_{4} &= -\frac{1}{17}\lp -7+\sqrt{49+r_{1}}\rp \\
	a_{5} &=-\frac{1}{17}\lp -7-\sqrt{49+r_{2}}\rp  \\
	a_{6} &= -\frac{1}{17}\lp -7+\sqrt{49+r_{2}}\rp \\
	a_{7} &= -\frac{1}{17}\lp -7-\sqrt{49+r_{3}}\rp \\
	a_{8} &= -\frac{1}{17}\lp -7+\sqrt{49+r_{3}}\rp 
  \end{split}
\quad \quad \quad \quad
  \begin{split}
    b_{1} &=14/17\\
		b_{2}  &= 12/17\\
		b_{3} &= 10/17\\
		b_{4}  &= 8/17\\
		b_{5} &= 6/17\\
		b_{6} &= 4/17\\
		b_{7} &= 2/17\\
		b_{8}  &= 0
  \end{split}
\end{equation}
where $r_{1},r_{2},r_{3}$ were defined in \nref{rs}. 

As before, we notice that a field redefinition of the form $\tilde P = z^{3/17} p$ will remove the terms with no derivatives. This time we get eight solutions for $\partial_{z}p$, seven $_7F_7$ functions and one Meijer-G function, the latter of which we again discard. As in the previous case, we take a linear combination of the seven hypergeometric function to eliminate polynomial terms scaling as $z^{-a_{m}}$ in the large $z$ expansion. The final result is as written in \nref{p7sol}.

In this case the expressions for the coefficients are sufficiently complicated that we do not bother trying to write their exact expressions. Their decimal approximations are
\begin{equation}	
\begin{split}\la{coeff}
	v_{0} &=9.379770542\\
	v_{1} &=-44.56762638\\
	v_{2} &=112.2906674\\
	v_{3} &=-180.1624337\\
	v_{4} &=196.2434578\\
	v_{5} &=-148.5448127\\
	v_{6} &=77.97782018\\
	v_{7} &=-26.70776367
\end{split}
\end{equation}


\section{Plots of $\tilde P(z)$}\la{Ptildeplots}
In this Appendix we include plots of the functions $\tilde P(z)$ themselves. From these plots we can see that in all three cases, $\tilde P$ has support in a range where $z$ is order one. We plot as a function of $z^{\frac{2}{3+2a}}$, which is the variable linear in $E$. 
 \begin{center}
     \includegraphics[width=0.7\textwidth]{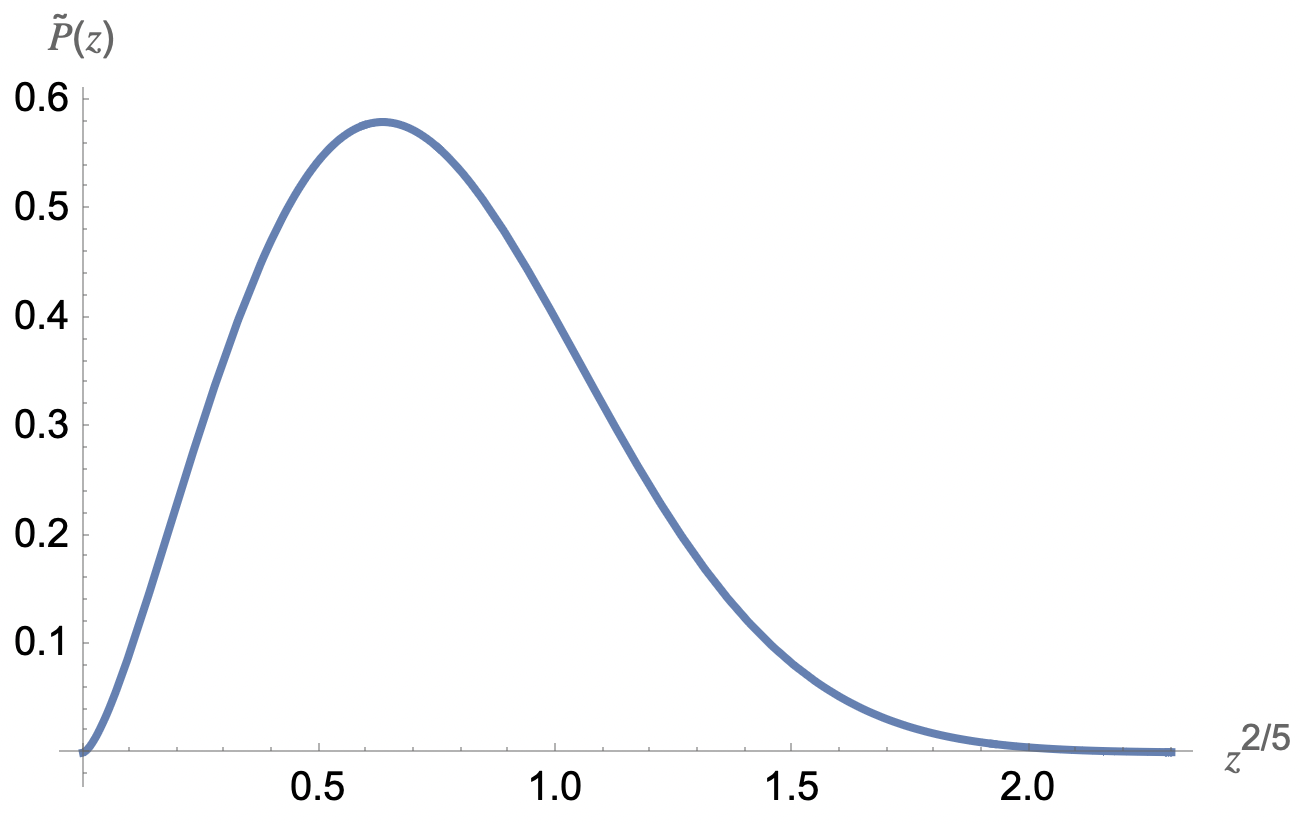}
    
    \textbf{(a)}
    
  \includegraphics[width=0.7\textwidth]{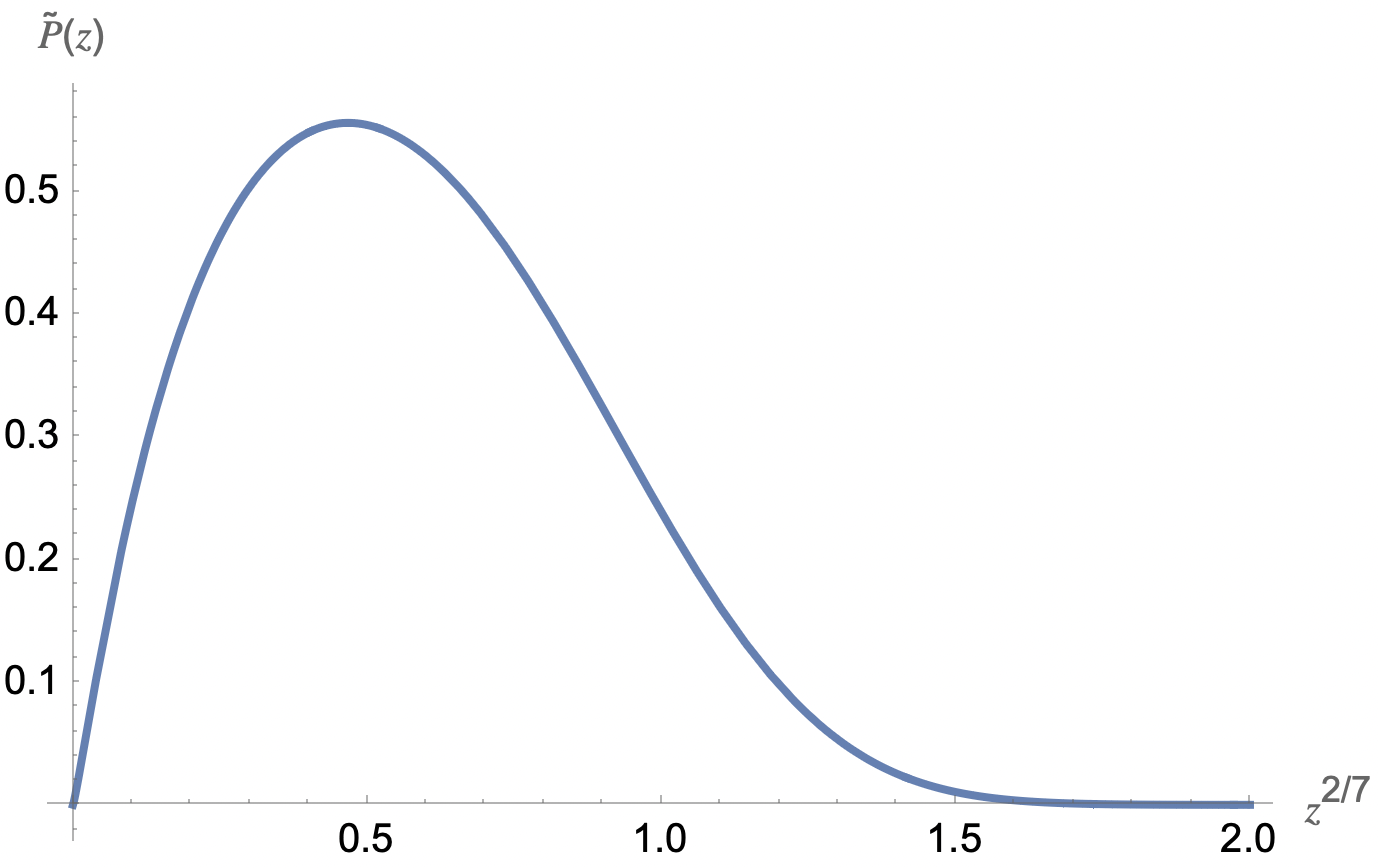}
  
  \textbf{(b)}
  
    \includegraphics[width=0.7\textwidth]{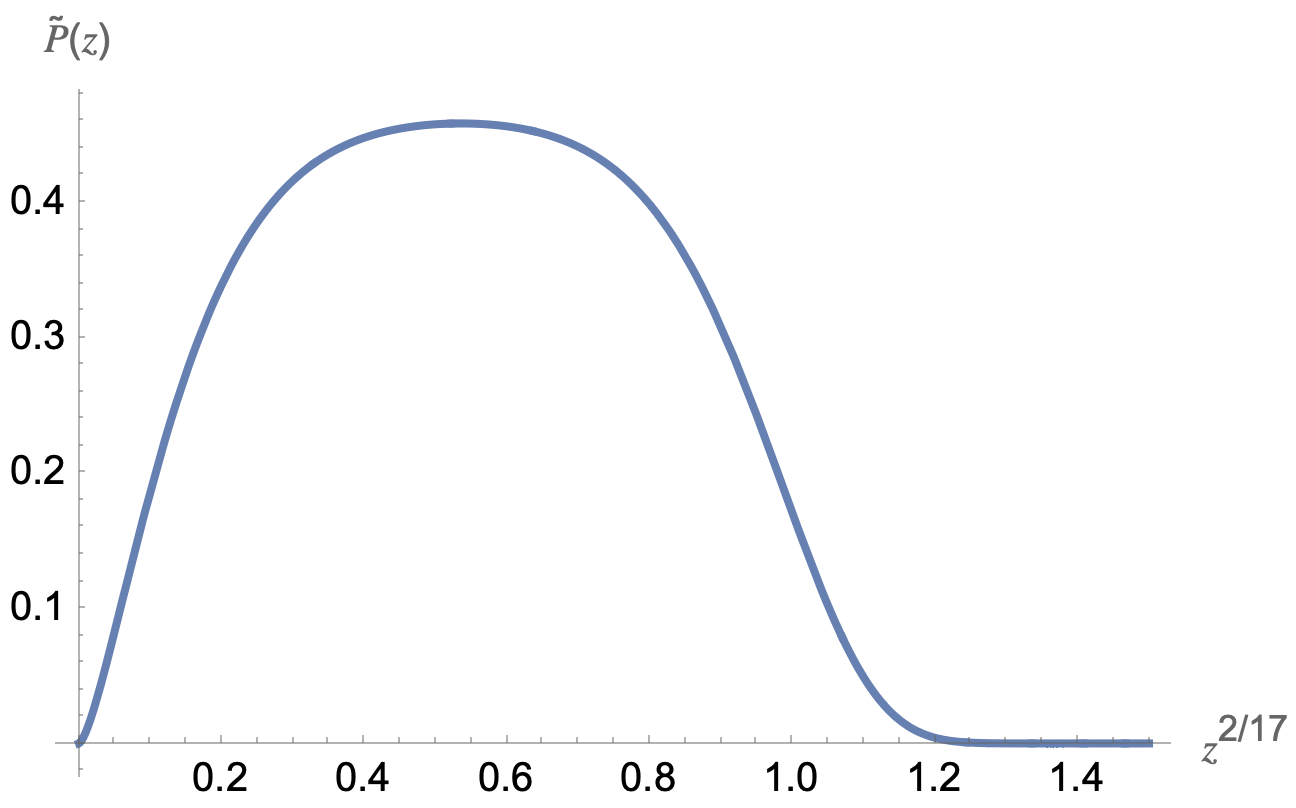}
    
    \textbf{(c)}
    
     \captionof{figure}{Plots of the three functions $\tilde P(z)$ which make up the attractor solutions. a) Scalar emission, see \nref{tildePa1}.  b) Photon emission, see \nref{tildePa3}.  c) Di-photon emission, see \nref{p7sol} and \nref{coeff}.}
    
 	 \end{center}

\section{Derivation of absorption cross section in the effective theory}\la{sigmaabs}

In this appendix we calculate the absorption cross section for a neutral massless scalar in the effective theory \nref{Hint}. 

Suppose $\phi$ begins in a coherent state $|\alpha\rangle$ with frequency $\omega$,
\begin{align}
	a_{\omega}|\alpha \rangle = \alpha |\alpha \rangle
\end{align}
The expected particle number at $t = 0$ is
\begin{align}\la{Nw}
	\langle \alpha | a_{\omega}^{\dagger}a_{\omega}|\alpha \rangle =|\alpha|^{2}
\end{align}
We would like to know how the particle number changes after the field has interacted with the black hole for some time $T$. This tells us how much of the field has been absorbed. Since the black hole and scalar field are weakly coupled, we simply time evolve the number operator and expand to quadratic order in $H_{\text{int}}$. We have
\begin{align}\nonumber
	a_{\omega}(T) = U(T)^{\dagger}a_{\omega} U(T) = a_{\omega} + i \int_{0}^{T} dt [H(t),a_{\omega}] -\int_{0}^{T} dt \int_{0}^{T} dt'[H(t'),[H(t),a_{\omega}]]\theta(t-t')+...
\end{align}
and a similar expression for $a_{\omega}^{\dagger}$. Plugging $a(T)$ and $a^{\dagger}(T)$ into \nref{Nw},
	\begin{align}\la{aexp}
	&\langle \alpha | a_{\omega}^{\dagger}(T) a_{\omega}(T)|\alpha\rangle -\langle \alpha | a_{\omega}^{\dagger}a_{\omega}|\alpha \rangle = -  \int dt \int dt' \langle \alpha |[H(t), a_{\omega}^{\dagger}][H(t'),a_{\omega}]|\alpha \rangle \\
	&-\int dt \int dt' \theta(t-t')\langle \alpha | a^{\dagger}_{\omega}[H(t'),[H(t),a_{\omega}]]|\alpha\rangle - \int dt \int dt'  \theta(t-t') \langle \alpha | [H(t'),[H(t),a_{\omega}^{\dagger}]]a_{\omega}|\alpha\rangle +...\nonumber
\end{align}
where we have dropped terms linear in $H$ under the assumption that $\langle O \rangle = 0$. 
Note that the first term in \nref{aexp} would be nonzero even in vacuum, while the second two would not. This is because the first term describes spontaneous emission. Evaluating the commutators,
\begin{align}
	-  \int dt \int dt' \langle \alpha |[H(t), a^{\dagger}][H(t'),a]|\alpha \rangle  &= \frac{g^{2} T}{2\omega}\int_{-T}^{T} dt e^{-i \omega t}\langle O(t) O(0)\rangle 
\end{align}
This gives a positive contribution to the particle number and represents the probability that the black hole spontaneously emits a mode of frequency $\omega$ in the time $t$, as shown previously in section \ref{ETemission}. This probability is independent of $|\alpha|^{2}$, as expected. 

The second two terms together describe absorption and stimulated emission. These contributions are quadratic in $\alpha$ because the double commutators are linear in the $a_{\omega}$s and $a_{\omega}^{\dagger}$s, which paired with the second $a_{\omega}/a_{\omega}^{\dagger}$ insertion produces a number operator. In all, we find
\begin{equation}\la{refprob}
\begin{split}
	\langle \alpha | a_{\omega}^{\dagger}(T) a_{\omega}(T)|\alpha\rangle -|\alpha|^{2} &=\frac{g^{2} T}{2\omega}\int_{-T}^{T} dt e^{-i \omega t}\langle O(t) O(0)\rangle  -\frac{g^{2}|\alpha|^{2}T}{2\omega} \int_{-T}^{T} dt e^{i \omega t}   \langle [O(t),O(0)]\rangle \\
	&=T \lp (|\alpha|^{2}+1)\frac{g^2}{2 \omega}G(-\omega) -\frac{g^{2}}{2 \omega}|\alpha|^{2}G(\omega) \rp 
	\end{split}
\end{equation}

In evaluating \nref{refprob} we have dropped a term which tends to zero in the $T \to \infty$ limit. We see that the expected particle number is depleted by absorption, which is the term proportional to $G(\omega)$, and increased by a combination of spontaneous and stimulated emission, which is the term involving $G(-\omega)$. The latter is proportional to $|\alpha|^{2} + 1$, as in atomic physics. 

The expected number of particles absorbed per unit time is the difference between the flux of ingoing and the flux of outgoing particles:
\begin{align}\la{fluxdiff}
	\Phi_{\text{in}} - \Phi_{\text{out}} = -\frac{g^{2}}{2 \omega}G(-\omega) + \frac{g^{2}|\alpha|^{2}}{2 \omega} \lp G(\omega) - G(-\omega) \rp 
\end{align}

\bibliographystyle{apsrev4-1long}
\bibliography{GeneralBibliography.bib}

	\end{document}